\documentclass[aps,pre,showpacs,amsmath,amssymb,amsfonts,superscriptaddress,lengthcheck,twocolumn,longbibliography]{revtex4-1}
\pdfoutput=1
\usepackage{color}
\usepackage{graphicx}
\usepackage{amsthm}
\usepackage{subfigure}
\usepackage{tabu}
\usepackage{dcolumn}
\usepackage{bm}
\usepackage{epsf}
\usepackage{color}
\usepackage[colorlinks=true,citecolor=blue,linkcolor=blue,urlcolor=blue]{hyperref}
\usepackage{hhline}

\newcommand{\be}{\begin{equation}}
\newcommand{\ee}{\end{equation}}

\newtheorem{theorem}{Theorem}

\begin{document}

\title{Compatibility of linear-response theory with the Second Law of Thermodynamics and the emergence of negative entropy production rates}

\author{Pierre Naz\'e}
\email[]{p.naze@ifi.unicamp.br}
\affiliation{Instituto de F\'isica `Gleb Wataghin', Universidade Estadual de Campinas, 13083-859, Campinas, S\~{a}o Paulo, Brazil}

\author{Marcus V. S. Bonan\c{c}a}
\email[]{mbonanca@ifi.unicamp.br}
\affiliation{Instituto de F\'isica `Gleb Wataghin', Universidade Estadual de Campinas, 13083-859, Campinas, S\~{a}o Paulo, Brazil}

\date{\today}

\begin{abstract}

The reliability of physical theories depends on whether they agree with well established physical laws. In this work, we address the compatibility of the Hamiltonian formulation of linear-response theory with the Second Law of Thermodynamics. In order to do so, we verify three complementary aspects often understood as statements of the Second Law: 1. No dissipation for quasistatic process; 2. Dissipation for finite-time processes; 3. Positive entropy production rate. Our analysis focus on two classes of nonequilibrium isothermal processes: slowly-varying and finite-time but weak ones. For the former, we show that these aspects are easily verified. For the later, we present conditions for the achievement of the first two aspects. We also show that the third one is not always verified, presenting an example based on Brownian motion in which we observe negative values in the entropy production rate. In particular, we compare linear-response and exact results for this example.

\end{abstract}

\maketitle

\section{Introduction}
\label{sec:intro} 

Linear-response theory is one of the most used frameworks to describe nonequilibrium statistical physics phenomena. Having notorious success in the calculation of kinetic coefficients in transport processes, such as electric conductivity \cite{mori1956, nakano1956}, dielectric relaxation \cite{debye1929, frohlich1949, bottcher1973, bordewijk1978} and nuclear magnetic susceptibility \cite{bloch1946, abragam1961, slichter1978, hashitsume1970}, the main idea underneath the theory is to provide the response of an observable due to a weak perturbation. The response is encoded by the so-called response function or, equivalently, the relaxation function \cite{kubo1985}. Additionally, a great advantage of linear-response theory is its twofold aspect of describing a system either by using a microscopical approach or a phenomenological one. In the former case, the relaxation function is deduced directly by solving the Hamiltonian equations (or the Heisenberg ones in the quantum case). In the phenomenological approach, although the underlying Hamiltonian formalism is the same, the relaxation function is obtained from the experimental measurement of the response. In this case, it is quite natural to expect that such relaxation function must be very related to the thermodynamic aspects of the system. However, it is not very clear yet whether the linear-response results derived either from microscopic or phenomenological inputs are entirely compatible with thermodynamics and how this compatibility takes place, although some attempts have been made in that sense \cite{grigolini1995}. In the present work, we try to fill this gap, finding mathematical conditions on the relaxation function that make the linear-response theory compatible with the Second Law of Thermodynamics. Since we are focusing on the Hamiltonian formulation of linear-response theory, our task is related to the old problem of understanding how the macroscopic phenomena are related to the microscopic laws of motion \cite{kemble1939,ruelle1993,lebowitz1993,penrose2005,lebowitz2007}.

In our analysis, we will consider driven classical systems in the presence of a heat bath and three aspects often taken as statements of the Second Law will be considered: the first one says that there is no dissipation for systems driven by quasistatic process; the second one says that there is dissipation for systems driven by finite-time process; the third one corresponds to the positivity of the entropy production rate. Considering the two classes of nonequilibrium processes described by linear-response theory, namely, slowly-varying and finite-time but weak ones, each one of the aspects just mentioned will be verified, either by finding mathematical conditions on the relaxation function or by presenting a counterexample. This will be the case when we analyze the third aspect for finite-time but weak processes. There we will find that linear-response theory predicts negative values of entropy production rates for the paradigmatic example of driven Brownian motion.

The proposition of the positivity of the entropy production rate as an equivalent statement of the Second Law of Thermodynamics dates back to the mid-twentieth century in the works of Prigogine and contemporaries \cite{mazur1962, prigogine1967}. In their local formulation of the Second Law, it is stated that the differential of the internal entropy production of the system must be positive, which would imply that the entropy production rate has to be positive as well. In that manner, Prigogine recovers well-established results and since then the positivity of the entropy production rate has been used either as a premisse or as a goal to be achieved \cite{schnakenberg1976,mcadory1977,spohn1978,lebowitz1978,alicki1979,cohen1995,ruelle1996,seifert2005}. However, the existence of negative entropy production rates has become a topic of intense research mainly because of its supposed relation with non-Markovian aspects of the dynamics of open quantum systems \cite{plenio2014, breuer2016, alonso2017}. Although a considerable amount of examples has been presented in the last years trying to establish a connection between negative rates and non-Markovianity \cite{maniscalco2016, pati2017, marcantoni2017, campbell2018, feng2018}, our understanding about it is still improving \cite{majumdar2019, esposito2018}.

This work is organized in the following form: in Sec.~\ref{sec:c2lt} we show how to connect linear-response theory to the three aspects of the Second Law mentioned previously and address the nonequilibrium regions where our analysis will be done; after that, in Sec.~\ref{sec:lrt}, we review the main elements of linear-response theory that are necessary for the development of this work; in Secs.~\ref{sec:csvp} and \ref{sec:cftw}, we address the above-mentioned compatibility for slowly-varying and finite-time but weak processes. In particular, Sec.~\ref{subsec:nepr} presents an example where we observe negative values of the entropy production rate. We make our final remarks in Sec.~\ref{sec:final}.

\section{Connecting linear-response theory to the 2nd Law}
\label{sec:c2lt}

According to the Second Law of Thermodynamics, the entropy variation of a total isolated system, after we have changed a control parameter $\lambda$ from $\lambda_{0}$ to $\lambda_{0}+\delta\lambda$ (we will restrict our analysis to a single control parameter) in a time interval $\tau$, is positive or zero,
\be
\Delta S_{\text{tot}} \ge 0.
\label{eq:secondlaw}
\ee
To connect linear-response theory with the Second Law, it is convenient to express (\ref{eq:secondlaw}) in a more suitable way. To this end, we restrict ourselves to the situation in which the total system is composed of a system of interest (or simply system) coupled to a heat bath at temperature $T$. Considering $\Delta S$ and $\Delta S_B$ respectively as the entropy variation of the system and the heat bath, $Q$ as the average heat received by the system, $W$ as the average work performed on the system by the external agent and $\Delta F$ as the variation of free energy between the final and initial equilibrium state of the system, where $F = U-TS$, we have
\begin{align*}
\Delta S_{\text{tot}} = \Delta S + \Delta S_B  \quad & \Rightarrow \quad \Delta S_{\text{tot}} = \Delta S - Q/T\\
\quad & \Rightarrow \quad T\Delta S_{\text{tot}} = T\Delta S + W - \Delta U\\
\quad & \Rightarrow \quad T\Delta S_{\text{tot}} = W-\Delta F,
\end{align*}
where we used the Clausius theorem and the First Law of Thermodynamics, $\Delta U=Q+W$, in the first and third implication respectively. Defining the irreversible work $W_{\text{irr}}$ as
\be
W_{\text{irr}} = W-\Delta F,
\label{eq:wirrwf}
\ee
we have
\be
T\Delta S_{\text{tot}} = W_{\text{irr}}.
\ee
Thus, the irreversible work $W_{\text{irr}}$ can be used as a measure of entropy production $\Delta S_{\text{tot}}$ (see for instance Ref. \cite{crooks1999}). 

In the last decades, expressions for $W_{\mathrm{irr}}$ have been derived using linear-response theory for the purpose of finding optimal finite-time processes \cite{mou1994,antonelli1997,crooks2012,deffner2014,bonanca2015,bonanca2018}. The connection between linear-response theory and the Second Law can be established then through these expressions for $W_{\mathrm{irr}}$ which are asked to verify the following statements:
\begin{enumerate}
\item No dissipation for quasistatic process:
\begin{equation}
\lim_{\tau\rightarrow\infty}W_{\text{irr}}(\tau) = 0,
\label{eq:qstatic}
\end{equation} 
\item Dissipation for finite-time processes:
\begin{equation}
W_{\text{irr}}(\tau) \ge 0,
\label{eq:diss}
\end{equation} 
\item Positive entropy production rate:
\begin{equation}
\dot{W}_{\text{irr}}(t) \ge 0.
\end{equation} 
\end{enumerate}
It is worth clarifying that processes in which the system reaches a stationary state with non-vanishing currents for fixed values of the control parameter $\lambda$ are ruled out of our considerations. 

\begin{figure}
    \includegraphics[scale=0.42]{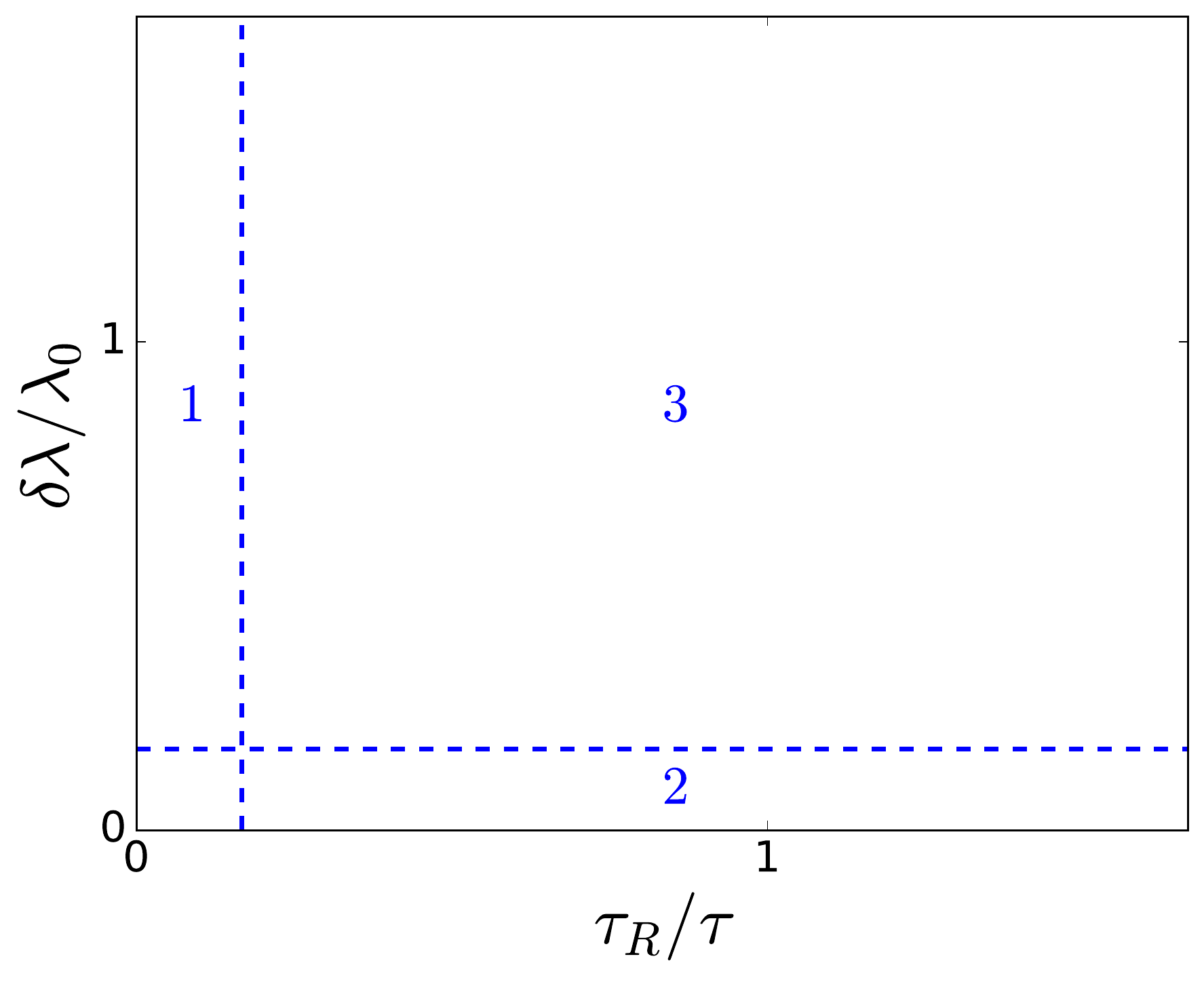}
    \caption{(Color online) Diagram of nonequilibrium regions. Region 1: slowly-varying processes, Region 2: finite-time but weak processes and Region 3: arbitrarily far from equilibirum processes.}
\label{fig:diagram_noneq}
\end{figure}

Our analysis will focus on two nonequilibrium regimes represented in Fig.~\ref{fig:diagram_noneq} by regions 1 and 2. To each of them corresponds a different class of nonequilibrium processes. Slowly-varying processes belong to region 1, where the duration of the process $\tau$ is large compared to the system's relaxation time $\tau_{R}$, but the variation $\delta\lambda$ is arbitrary. In such regime, the expression for $W_{\mathrm{irr}}$ deduced by means of linear-response theory is given by \cite{mou1994,antonelli1997,crooks2012,deffner2014}
\be
W_{\text{irr}} = \beta\int_0^\tau\dot{\lambda}^2(t)\tau_R[\lambda(t)] \chi[\lambda(t)]dt,
\label{eq:wirrsvp}
\ee
where $\beta=(k_B T)^{-1}$, $k_{B}$ is Boltzmann constant and $\dot{\lambda}$ is the time derivative of the protocol $\lambda(t)$. We denote by $\tau_R[\lambda(t)]$ and $\chi[\lambda(t)]$ the parametric variation of the system's relaxation time and of the fluctuations of the generalized force conjugated to $\lambda$ respectively (in the case of a gas, if the volume is the control parameter then the pressure is the generalized force conjugated to it). We give more precise definitions of these quantities in the following Sections (see Eqs.~(\ref{eq:variance}) and (\ref{eq:timerelax}) of Sec.~\ref{sec:csvp}).

In region 2, we have finite-time but weak processes. In this regime, the relative change $\delta\lambda/\lambda_{0}$ must be small but the duration $\tau$ of the process can be arbitrary. In this case, using linear-response theory, the irreversible work expression reads \cite{bonanca2015}
\begin{equation}
W_{\text{irr}}(\tau) = \int_0^\tau\int_0^{t}\Psi_0(t-t')\dot{\lambda}(t')\dot{\lambda}(t)dt'dt,
\label{eq:irrwork}
\end{equation}
where $\Psi_{0}(t)$ is the so-called relaxation function \cite{kubo1985} (see its definition in Eq.~(\ref{eq:relax}) of Sec.~\ref{sec:lrt}). Last but not least, region 3 represents processes arbitrarily far from equilibrium, in which both perturbation and duration of the process are arbitrarily chosen outside near-equilibrium regions, and where linear-response theory does not hold anymore. Our goal is to find mathematical and physical conditions under which the functionals (\ref{eq:wirrsvp}) and (\ref{eq:irrwork}) and the quantities appearing in them lead to an agreement with the three aspects of the Second Law mentioned above. Succint deductions of such irreversible work expressions can be found in Appendix \ref{app:b}. For more details, see Refs.~\cite{deffner2014,bonanca2015}.

It is worth emphasizing that $W_{\mathrm{irr}}$ is an average over several microscopic realizations of the protocol $\lambda(t)$, each one furnishing a different value of work $\mathcal{W}$. Hence, $\mathcal{W}_{\mathrm{irr}}\equiv \mathcal{W}-\Delta F$ is a fluctuating quantity that obeys very well-known fluctuation theorems, namely, the Jarzynski equality \cite{jarzynski1997}
\begin{equation}
\left\langle e^{-\beta(\mathcal{W}-\Delta F)}\right\rangle = \left\langle e^{-\beta\mathcal{W}_{\mathrm{irr}}}\right\rangle = 1\,,
\label{eq:jarzynski}
\end{equation}
and Crooks fluctuation theorem \cite{crooks1999}
\begin{equation}
\frac{P_{F}(\mathcal{W})}{P_{R}(-\mathcal{W})} = \frac{P_{F}(\mathcal{W_{\mathrm{irr}}})}{P_{R}(-\mathcal{W}_{\mathrm{irr}})} = e^{\beta(\mathcal{W}-\Delta F)} = e^{\beta \mathcal{W}_{\mathrm{irr}}}\,,
\end{equation}
where $P_{F}$ and $P_{R}$ denote the usual work distributions obtained after several realizations of the forward (described by $\lambda(t)$) and time-reversed protocols.

We show in Sec.~\ref{subsec:nepr} by means of stochastic thermodynamics and linear-response theory that, along the process induced by   $\lambda(t)$, the instantaneous rate $\dot{W}_{\mathrm{irr}}$ can assume \emph{negative} values for the paradigmatic example of driven Brownian motion. This seems to be in contrast to previous definitions of entropy production rates in stochastic thermodynamics \cite{seifert2005}. We will postpone to Sec.~\ref{sec:final} additional comments about this point. In any case, we would like to stress that the inequality $\langle \mathcal{W}_{\mathrm{irr}}\rangle = W_{\mathrm{irr}}\geq 0$, derived from the integral fluctuation theorem (\ref{eq:jarzynski}), does not imply $\dot{W}_{\mathrm{irr}}\geq 0$.

\section{Hamiltonian formulation of linear-response theory}
\label{sec:lrt}

In this section, we review the main elements of linear-response theory that we find necessary for our purposes. As mentioned in the previous section, we consider a system coupled to a heat bath at temperature $T$ that is driven out of equilibrium by the switch of a certain control parameter $\lambda$. We denote this by
\be
\lambda(t)=\lambda_0+g(t)\delta\lambda,
\label{eq:protocol1}
\ee
with $g(0)=0$ and $g(\tau)=1$, where $\tau$ is the duration of the process. In a Hamiltonian approach, this means that the time-dependent Hamiltonian $\mathcal{H}(\lambda(t))$ of system plus heat bath is driven from $\mathcal{H}(\lambda_0)$ to $\mathcal{H}(\lambda_0+\delta\lambda)$ in a time interval $\tau$ by some external agent, according to the protocol $\lambda(t)$ (or, equivalently, $g(t)$). We will focus here on the calculation of the work performed on the system since, as explained in Sec.~\ref{sec:c2lt}, it will be the necessary link to connect linear-response theory to the Second Law of Thermodynamics. Our starting point is the following expression for the work performed on the system,
\be
W = \int_0^\tau dt\,\dot{\lambda}(t)\,\overline{\partial_\lambda \mathcal{H}}(t)\,,
\label{eq:1stlaw}
\ee
where $\dot{\lambda}:=d\lambda/dt$, $\overline{A}$ denotes the nonequilibrium average of the observable $A$ and the quantity $\overline{\partial_{\lambda} \mathcal{H}}:=\overline{\partial\mathcal{H}/\partial\lambda}$ is the generalized force. We are concerned with the linear response of the generalized force due to the variation of the parameter $\lambda$. Therefore, we assume that the system is weakly perturbed, that is, $\delta\lambda/\lambda_0 \ll 1$. Using linear-response theory (see Appendix \ref{app:a} for more details), the expression for the generalized force up to first order in $\delta\lambda$ reads
\begin{eqnarray}
\lefteqn{\overline{\partial_\lambda \mathcal{H}}(t) =}\nonumber\\
 &&\left\langle\partial_\lambda \mathcal{H}\right\rangle_0-\delta\lambda \Theta_0 g(t) +\delta\lambda\int_{0}^{t} du\Psi_0(u)\frac{dg}{dt'}\Bigr|_{t'=t-u},
 \label{eq:genforce}
\end{eqnarray}
where $\langle ... \rangle_0$ is an equilibrium average taken on the initial canonical distribution, $\exp{(-\beta \mathcal{H}(\lambda_{0}))}/Z(\beta,\lambda_{0})$ ($Z(\beta,\lambda_{0})$ being the partition function) and $\Psi_0(t)$ is the relaxation function, given, in our specific case, by 
\be
\Psi_0(t) = \beta\left\langle\partial_\lambda \mathcal{H}(q_0,p_0)\partial_\lambda \mathcal{H}(q_t,p_t)\right\rangle_0-\mathcal{C}\,,
\label{eq:relax}
\ee
after using Kubo formula \cite{kubo1985} (see Eqs.~(\ref{eq:a9}) to (\ref{eq:a10a}) in Appendix~\ref{app:a}). We denote by $(q_t,p_t)$ a phase-space point of the entire system at the instant $t$ and $\mathcal{C}$ is a constant defined by \cite{kubo1985}
\be
\mathcal{C} = \beta\lim_{s\rightarrow 0} s\int_0^\infty e^{-st}\left\langle\partial_\lambda \mathcal{H}(q_0,p_0)\partial_\lambda \mathcal{H}(q_t,p_t)\right\rangle_0 dt,
\label{eq:kuboconst}
\ee
whose purpose is to guarantee that the system attains the correct new equilibrium state (it is assumed then from the very beginning that the auto-correlation function in Eq. (\ref{eq:relax}) decays due to the interaction with the heat bath). The purpose of the subscript ``$0$" is to emphasize that the initial equilibrium averages were taken with $\lambda=\lambda_{0}$. Finally, the constant $\Theta_0$ is defined as (see Appendix \ref{app:a})
\be
\Theta_0 := \Psi_0(0)-\left\langle\partial^2_{\lambda\lambda} \mathcal{H}\right\rangle_0.
\ee

To calculate the work performed on the system, it is possible to use Eq.~(\ref{eq:genforce}) in two different regimes characterized previously by regions 1 and 2 of Fig.~\ref{fig:diagram_noneq} \cite{mou1994,antonelli1997,crooks2012,deffner2014,bonanca2015,bonanca2018} (see Appendix~\ref{app:b}). In both cases, the relaxation function $\Psi_{0}(t)$ is the central object of the theory and its exact expression demands the solutions of Hamilton's equations, which is not a very easy task to be accomplished. In order to circumvent this problem, the relaxation function can alternatively be modeled using phenomenological information \cite{kubo1985,kubo1972}. Although this way of obtaining $\Psi_{0}(t)$ is never in full agreement with the underlying Hamiltonian dynamics, it can be made approximately consistent with it. This is achieved through the so-called sum rules, which are constraints that the phenomenological relaxation function must satisfy to match its expected microscopic requirements \cite{kubo1985,kubo1972} (see Appendix~\ref{app:c} for more details). For instance, Hamiltonian dynamics demand Eq.~(\ref{eq:relax}) to be even under the change $t\to -t$. Thus, this constraint must be imposed to any phenomenological expression intended to be used as a valid relaxation function.

As it will be shown in Sec.~\ref{sec:cftw}, this parity property will be very important to make the linear-response expression of Eq.~(\ref{eq:1stlaw}) compatible with the Second Law of Thermodynamics. Nevertheless, we want to emphasize that $\Psi_{0}(-t)=\Psi_{0}(t)$ holds in the particular case of interest here of nonequilibrium work due the variation of a \emph{single} external parameter. Different parities are indeed possible for more general cases \cite{kubo1985,kubo1972}.

\section{Compatibility for slowly-varying processes}
\label{sec:csvp}

The irreversible work for slowly-varying processes is given by Eq.~(\ref{eq:wirrsvp}) which depends on two quantities we have not defined properly yet. The first one is the variance $\chi$ of the generalized force defined as
\be
\chi(\lambda_0) = \langle\left(\partial_{\lambda} \mathcal{H}\right)^{2}\rangle_{0}- \left\langle\partial_{\lambda} \mathcal{H}\right\rangle^{2}_{0},
\label{eq:variance}
\ee
where the dependence in $\lambda_0$ emphasizes that the canonical average is taken over the initial equilibrium distribution. The second quantity is the relaxation time defined in linear-response theory by 
\be
\tau_R(\lambda_0) = \int_{0}^\infty \frac{\Psi_0(t)}{\Psi_0(0)}dt,
\label{eq:timerelax}
\ee
where the dependence in $\lambda_0$ emphasizes again that averages were taken with the initial equilibrium distribution. Observing Eq.~(\ref{eq:wirrsvp}), one concludes that $W_{\text{irr}}$ is given in terms of equilibrium quantities that vary parametricaly in time through the protocol $\lambda(t)$. This suggests that, in this regime, the system slightly deviates from a sequence of equilibrium states as $\lambda$ changes in time. In other words, the $\chi[\lambda(t)]$ and $\tau_R[\lambda(t)]$ used in Eq.~(\ref{eq:wirrsvp}) are still given by Eqs.~(\ref{eq:variance}) and (\ref{eq:timerelax}) evaluated though at the value $\lambda(t)$ instead of $\lambda_0$. Additionally, it only makes sense to consider relaxation functions leading to positive and finite relaxation times.

\subsection{No dissipation for quasistatic processes}

Defining $u=t/\tau$, we can rewrite Eq.(\ref{eq:wirrsvp}) as
\be
W_{\text{irr}} = \frac{\beta}{\tau}\int_0^1\dot{\lambda}^2(u)\tau_R[\lambda(u)] \chi[\lambda(u)]du.
\label{eq:wirrsv}
\ee
We note that the dependence on the switching time occurs only in the factor $1/\tau$ \cite{deffner2014}. This is so because $\lambda(t) = \lambda_{0} + \delta\lambda g(t)$ satisfies the boundary conditions $\lambda(0) = \lambda_{0}$ and $\lambda(\tau)=\lambda_{0}+\delta\lambda$, implying that $g(t)$ is indeed a function of $t/\tau$. In this manner, we have immediately
\be
\lim_{\tau\rightarrow\infty} W_{\text{irr}} = \lim_{\tau\rightarrow\infty} \frac{\beta}{\tau}\int_0^1\dot{\lambda}^2(u)\tau_R[\lambda(u)] \chi[\lambda(u)]du = 0.
\ee
\subsection{Dissipation for finite-time processes}

As the integrand of Eq. (\ref{eq:wirrsv}) is composed of positive functions, the integral leads to a positive result as well. Therefore, for a finite switching time $\tau$, we have
\be
W_{\text{irr}} = \frac{\beta}{\tau}\int_0^1\dot{\lambda}^2(u)\tau_R[\lambda(u)] \chi[\lambda(u)]du> 0.
\ee

\subsection{Positive entropy production rate}

Considering again Eq. (\ref{eq:wirrsvp}), the time derivative of $W_{\text{irr}}$ is composed of positive functions only. Thus,
\be
\dot{W}_{\text{irr}} = \beta\dot{\lambda}^2(t)\tau_R[\lambda(t)] \chi[\lambda(t)]>0.
\label{eq:powersv}
\ee

In addition, by analogy with the Thermodynamics of Linear Irreversible Processes \cite{mazur1962,prigogine1967}, we will define some nomenclature that will be useful for our discussion in what follows. Consider the factorization

\be
\dot{W}_{\text{irr}} = \dot{\lambda}(t)\left(\beta\tau_R[\lambda(t)] \chi[\lambda(t)]\dot{\lambda}(t)\right) = \dot{\lambda}(t) \mathcal{F}[\lambda(t)].
\label{eq:powersv2}
\ee
We call $\dot{\lambda}$ and $\mathcal{F}$ respectively the {\it affinity} and the associated {\it flux}. We observe that, in the situation where the process is driven externally, the affinity is related to how the external agent performs the process, namely, $\dot{\lambda}(t)$. In particular, for slowly-varying processes, the flux responds instantaneously to the affinity, which characterizes a {\it memoryless} process \cite{callen1985}.

\section{Compatibility for finite-time but weak processes}
\label{sec:cftw}

In the previous section, we have shown that the compatibility of the linear-response expression (\ref{eq:wirrsvp}) with the Second Law is straightforward and the only necessary requirement on the relaxation function is the convergence of the integral (\ref{eq:timerelax}) defining the relaxation time for all values of $\lambda$ between $\lambda_{0}$ and $\lambda_{0}+\delta\lambda$.
Our goal hereafter is to provide further constraints that any relaxation function must fulfill in order to achieve compatibility with the Second Law in region 2. As a first step, we will consider that the relaxation function $\Psi_{0}(t)$ has the following parity due to time-reversal symmetry,
\begin{equation}
\Psi_0(t) = \Psi_0(-t).
\label{eq:reflection}
\end{equation}
As we will see, such property will be essential to demonstrate the aspects that follow below. This is indeed a property inherited from the Hamiltonian definition of the relaxation function, which is an auto-correlation function calculated in the canonical equilibrium (see Eq.~(\ref{eq:relax})), and already discussed in Sec.~\ref{sec:lrt}.

\subsection{No dissipation for quasistatic processes}

We demonstrate now that Eq.~(\ref{eq:irrwork}) satisfies the first aspect of compatibility with the Second Law (see Eq.~(\ref{eq:qstatic})) assuming that the system thermalises with the heat bath. In other words, we assume as before that the relaxation time, defined by Eq.~(\ref{eq:timerelax}), is finite,
\begin{equation}
\int_0^\infty \frac{\Psi_0(t)}{\Psi_0(0)}dt < \infty.
\label{eq:finitetime}
\end{equation}
We observe first that we can rewrite the relaxation time as
\be
\tau_R = \frac{\tilde{\Psi}_0(0)}{\Psi_0(0)}, 
\ee
where $\tilde{\Psi}_0$ is the Laplace transform of $\Psi_0$. Thus, Eq.~(\ref{eq:finitetime}) leads to
\begin{align*}
\tilde{\Psi}_0(0) <\infty \quad &\Rightarrow\quad \lim_{s\rightarrow 0}\tilde{\Psi}_0(s)<\infty\\
&\Rightarrow\quad \lim_{s\rightarrow 0}s\tilde{\Psi}_0(s)=0\\
&\Rightarrow\quad \lim_{t\rightarrow \infty} \Psi_0(t) = 0,
\end{align*}
where in the last line we used the final value theorem. We remark that the reverse implication is not true as shown by the counterexample $\Psi_0(t)=\Psi_{0}(0)/(1+|t|)$. 

\begin{figure}
    \includegraphics[scale=0.42]{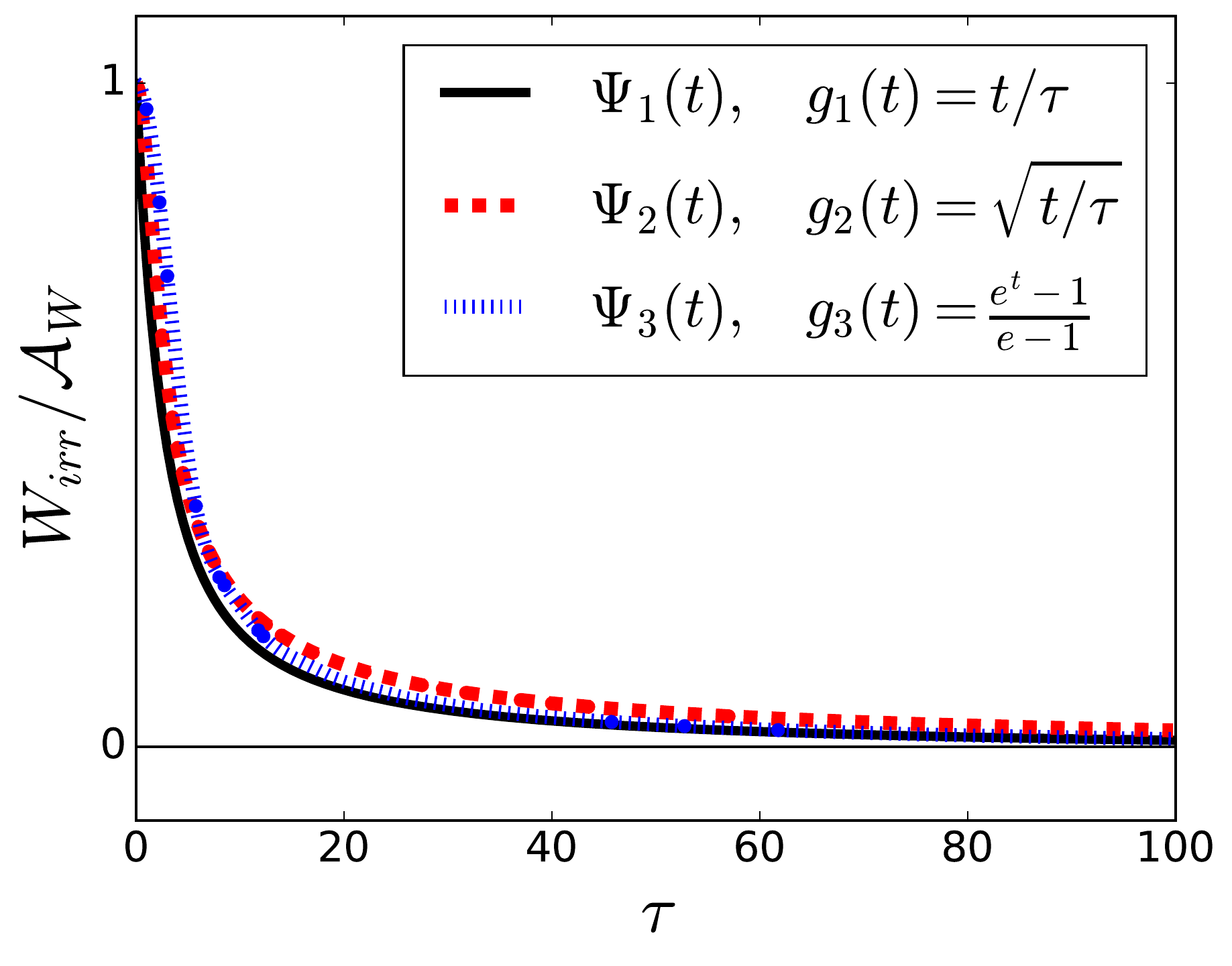}
    \caption{(Color online) Irreversible work given by Eq.~(\ref{eq:irrwork}) as a function of switching time $\tau$ for the relaxation functions defined in Eqs.~(\ref{eq:psis}) and protocols shown in the inset. In all cases we observe that dissipation decreases monotonically as the quasistatic limit is approached, $\tau \rightarrow \infty$. We chose the irreversible work unit $\mathcal{A}_W = \Psi_0(0){(\delta\lambda)}^2/2$}
\label{fig:qs_limit}
\end{figure}

Considering Eq.~(\ref{eq:reflection}) and the definitions $u=t/\tau$ and $v=t'/\tau$, we rewrite Eq.~(\ref{eq:irrwork}) as
\begin{equation}
W_{\text{irr}}(\tau) = \frac{1}{2}\int_0^1\int_0^1\Psi_0(\tau(u-v))\dot{\lambda}(u)\dot{\lambda}(v)dudv,
\label{eq:irrwork2}
\end{equation}
(for more details, see \cite{bonanca2015}). Therefore, we have
\begin{align*}
\lim_{\tau\rightarrow\infty}W_{\text{irr}}(\tau) &\propto \lim_{\tau\rightarrow\infty}\int_0^1\int_0^1\Psi_0(\tau(u-v))\dot{\lambda}(u)\dot{\lambda}(v)dudv \\
&= \int_0^1\int_0^1\left(\lim_{\tau\rightarrow\infty}\Psi_0(\tau(u-v))\right)\dot{\lambda}(u)\dot{\lambda}(v)dudv \\
&= 0,
\end{align*} 
in which we moved the limit inside the integral assuming that the integrand is well-behaved and used Eq.~(\ref{eq:reflection}) again. Figure \ref{fig:qs_limit} corroborates our result presenting the irreversible work given by Eq.~(\ref{eq:irrwork}) for different protocols and relaxation functions (the subscript ``0" was dropped for the sake of simplicity of notation)
\begin{subequations}\label{eq:psis}
\begin{align}
\Psi_1(t) &= \Psi_{1}(0)e^{-a_1|t|}, \label{eq:psi1}\\
\Psi_2(t) &= \Psi_{2}(0)e^{-a_2|t|}(\cos{(a_2 t)}+\sin{(a_2|t|)}),\label{eq:psi2}\\
\Psi_3(t) &= \Psi_{3}(0)J_0(a_3 t),\label{eq:psi3}
\end{align}
\end{subequations}
which satisfy Eqs.~(\ref{eq:reflection}) and (\ref{eq:finitetime}). We remark that $J_0$ is the Bessel function of the first kind for index $\alpha=0$ and $a_1$, $a_2$ and $a_3$ are positive free parameters. The relaxation functions $\Psi_1$ and $\Psi_2$ are commonly used to model respectively overdamped and underdamped Brownian motions \cite{kubo1985,deffner2014, bonanca2018} and $\Psi_3$ is motivated by correlation functions of spin systems \cite{glauber1963}.

It is worth emphasizing that there are no free parameters when the relaxation function is exactly obtained from the solutions of the equations of motion. Hence the expressions above represent possible phenomenological models for the behavior of the equilibrium correlation function given in Eq.~(\ref{eq:relax}). The free parameters of such models can be expressed in terms of thermodynamic quantities and parameters of the Hamiltonian through the expressions in Eq.~(\ref{eq:respocoef}) (for examples of how to do this, see Ref. \cite{kubo1985,deffner2014}).

\subsection{Dissipation for finite-time processes}
\label{sec:suddiss}

Dissipation for finite-time processes occurs if the following theorem is satisfied. 
\begin{theorem} 
Consider that the relaxation function is even with respect to the change $t\rightarrow -t$ due to time-reversal symmetry. The irreversible work given by Eq. (\ref{eq:irrwork}) is nonnegative if, and only if, the Fourier transform of the relaxation function is a nonnegative function
\begin{equation}
\hat{\Psi}_0(\omega) = \frac{1}{\sqrt{2\pi}}\int_{\mathbb{R}}e^{-i\omega t}\Psi_0(t)dt\ge 0.
\end{equation}
\end{theorem}
We will restrict ourselves to prove the first implication. The other one can be seen in detail in Ref.~\cite{feller2008} under the name {\it Bochner's theorem}. 
\begin{proof}
Firstly, $\hat{\Psi}_0$ is a real function since $\Psi_{0}(t)=\Psi_{0}(-t)$. Hence, if $\hat{\Psi}_0$ is a positive function, we have
\begin{align*}
W_{\text{irr}}[\dot{g}] &\propto \int_0^1\int_0^1 \Psi_0(\tau(u-v))\dot{\lambda}(u)\dot{\lambda}(v)dudv\\ 
&=\frac{1}{\sqrt{2\pi}}\int_0^1\int_0^1\int_{\mathbb{R}}e^{i\omega\tau (u-v)}\hat{\Psi}_0(\omega)\dot{\lambda}(u)\dot{\lambda}(v)d\omega dudv \\
&=\frac{1}{\sqrt{2\pi}}\int_{\mathbb{R}}\left|\int_0^1 e^{i\omega\tau u}\dot{\lambda}(u)du\right|^2 \hat{\Psi}_0(\omega)d\omega \ge 0.
\end{align*}
\end{proof}

We present below the Fourier transform of the relaxation functions defined in Eqs.~(\ref{eq:psis}), which satisfy the hypotheses of Theorem 1,
\begin{subequations}\label{eq:psishat}
\begin{align}
\hat{\Psi}_1(\omega) &= \Psi_{1}(0)\sqrt{\frac{2}{\pi}}\frac{a_1}{a_1^2+\omega^2}\,,\\
\hat{\Psi}_2(\omega) &= \Psi_{2}(0)\sqrt{\frac{2}{\pi}}\frac{4 a_2^3}{4a_2^4+\omega^4}\,,\\
\hat{\Psi}_3(\omega) &= \Psi_{3}(0)\sqrt{\frac{2}{\pi }}\frac{\theta (\omega/a_3 +1)-\theta (\omega/a_4 -1)}{\sqrt{a_3^2-\omega ^2}}\,,
\end{align}
\end{subequations}
where $\theta$ is the Heaviside step-function. We conclude that these relaxation functions are compatible with aspect (\ref{eq:diss}) the Second Law of Thermodynamics. In summary, phenomenological models of $\Psi_{0}(t)$ whose Fourier transform does not follow the conditions of Theorem 1 can violate aspect (\ref{eq:diss}) and therefore cannot be accepted.

A question that naturally arises is which are the properties on the relaxation function itself (i.e., not on its Fourier transform) that guarantee the non-negativity of the irreversible work. In fact, there is mathematical research currently investigating related issues \cite{tuck2006,peschanski2006,peschanski2014,peschanski2015,peschanski2016}. 

\subsection{Negative values in the entropy production rate}
\label{subsec:nepr}

Our last task is to verify the non-negativity of the entropy production rate as described by linear-response theory in region 2. Surprisingly, we will show that Eq.~(\ref{eq:irrwork}) predicts negative values of entropy production rate for a rather simple example. Our results also reveal how such negative values disappear as the process becomes ``memoryless", a notion that we clarify later on.

In what follows, we will compare our linear-response results with exact ones provided by stochastic thermodynamics \cite{seifert2012}. This comparison will show a nice agreement between both sets of results. Additionally, it will become clear how the linear-response expression for the entropy production rate clarifies the origin of negative values in contrast to the unappealing numerical solution furnished by stochastic thermodynamics.

Our example consists on a particle of mass $m$, immersed in a heat bath of temperature $T$ and subjected to a time-dependent harmonic potential with stiffness $\lambda(t)$. We model the dynamics of this particle through the following Langevin equation,
\be
m \ddot{x}+\gamma \dot{x} +\lambda(t) x = f(t),
\label{eq:langevin}
\ee
where $x(t)$ is the particle position at time $t$, $\gamma>0$ is the dissipation constant and $f(t)$ is a white noise which has the properties
\begin{subequations}\label{eq:corr_lang}
\begin{align}
\overline{\overline{f(t)}} &= 0,\\ 
\overline{\overline{f(t)f(t')}} &= 2\gamma k_B T \delta(t-t'),
\end{align}
\end{subequations}
where $\overline{\overline{(...)}}$ is the average taken over the noise history. We will consider the underdamped regime in which $\gamma/\omega_0 < 2$, with $\omega_0^2 =\lambda_0/m $.

Due to Eqs.~(\ref{eq:wirrwf}) and (\ref{eq:1stlaw}), the instantaneous power given by stochastic thermodynamics reads
\be
\dot{W}^{ST}_{\text{irr}}(t) = \frac{\dot{\lambda}(t)}{2}\left(\overline{\overline{x^2}}(t)-\frac{d F}{d\lambda}(\lambda(t))\right),
\label{eq:ipst}
\ee
where the term $dF/d\lambda$ is a generalized force that measures the rate of change of the equilibrium free energy along the process. In the underdamped regime, the mean squared displacement $\overline{\overline{x^2}}$ can be found solving the following system of equations \cite{seifert2008}
\begin{subequations}\label{eq:underdamped}
\begin{align}
\partial_t\overline{\overline{x^2}} &= 2 \overline{\overline{xp}}/m,\\
\partial_t\overline{\overline{p^2}} &= -2\lambda(t)\overline{\overline{xp}}-2\gamma \overline{\overline{p^2}}/m+2\gamma T,\\
\partial_t\overline{\overline{xp}} &= \overline{\overline{p^2}}/m-\lambda(t)\overline{\overline{x^2}}-\gamma\overline{\overline{xp}}/m.
\end{align}
\end{subequations}

According to Eq.~(\ref{eq:irrwork}), the linear-response expression for the instantaneous power in region 2 reads
\be
\dot{W}^{LR}_{\text{irr}}(t) = \dot{\lambda}(t)\left(\int_{0}^t\Psi_0(t-t')\dot{\lambda}(t')dt'\right),
\label{eq:iplr}
\ee
in which, again by analogy with the Thermodynamics of Linear Irreversible Processes, the first factor, $\dot{\lambda}(t)$, is the affinity and the second one the associated flux. In contrast to Eq.~(\ref{eq:powersv2}), the delayed response of the flux with respect to the affinity characterizes a process {\it with memory}. 

To obtain $\dot{W}_{\mathrm{irr}}^{LR}(t)$, we need a expression for the relaxation function $\Psi_{0}(t)$. According to its definition, Eq.~(\ref{eq:relax}), an exact expression demands the knowledge of the solutions of Hamilton's equations of particle plus heat bath. As explained in Sec.~\ref{sec:lrt}, to circumvent this problem we can use a phenomenological model that is minimally compatible with the Hamiltonian dynamics. In the present case, this can be done as follows: we first obtain $x(t)$ by solving Eq.~(\ref{eq:langevin}) for $\lambda(t)=\lambda_{0}$; then, since $\partial_{\lambda}\mathcal{H} = x^{2}/2$ (we are considering a Brownian particle in a harmonic trap), we plug into $\Psi_{0}(t)$ the following expression
\be
\Psi_0(t) = \frac{\beta}{4}\left\langle \overline{\overline{x^2}}(t)x^{2}(0)\right\rangle_{0}-\mathcal{C}\,,
\label{eq:relaxfunc}
\ee
according to Eqs.~(\ref{eq:relax}) and (\ref{eq:kuboconst}). It is clear that the phenomenological correlation function in Eq.~(\ref{eq:relaxfunc}) is obtained after taking two averages, namely, one over the noise history and another, a canonical one, over the initial conditions. The final analytical expression for $\Psi_{0}(t)$ reads
\be
\begin{split}
\Psi_0(t) = \frac{e^{-\gamma |t|}}{2 \beta  m^2 \text{$\omega_0 $}^4 \omega^2} \left[\left(2\omega_0^2-\omega^2\right) \cos \left(\omega t\right)\right.\\\left. -\gamma  \omega \sin \left(\omega |t|\right)-2 \text{$\omega_0 $}^2\right],
\label{eq:relax2}
\end{split}
\ee
where $\omega^2 = 4\omega_0^2-\gamma^2$. The absolute value of $t$ was added by hand as a minimal Hamiltonian requirement due to time-reversal symmetry (see the discussion in Sec.~\ref{sec:lrt}). We observe that Eq.~(\ref{eq:relax2}) leads to a finite relaxation time and a positive Fourier transform. Therefore, according to what was discussed in Sec.~\ref{sec:suddiss}, it is an acceptable relaxation function to model our system. 

The relaxation time obtained from Eqs.~(\ref{eq:timerelax}) and (\ref{eq:relax2}) is
\be
\tau_R = \frac{1}{2\gamma}+\frac{\gamma}{2\omega_0^2}\,.
\label{eq:taurr}
\ee
It is convenient to express $\gamma$ and $\omega_0$ in terms of the coupling between the system and heat bath, given by $\eta:=\gamma/\omega_0$, and the relaxation time (\ref{eq:taurr})
\be
\gamma = \frac{1+\eta^2}{2\tau_R},\quad \omega_0=\frac{1+\eta^2}{2\eta\tau_R}.
\ee
In particular, we have set $\tau_R=1$ in all the following results. The entropy production rates were computed using the protocol (\ref{eq:protocol1}), with
\be
g(t) = \frac{t}{\tau}+\sin{\left(\frac{2\pi t}{\tau}\right)}.
\label{eq:protocol}
\ee

\begin{figure*}[t]
    \subfigure{\includegraphics[scale=0.42]{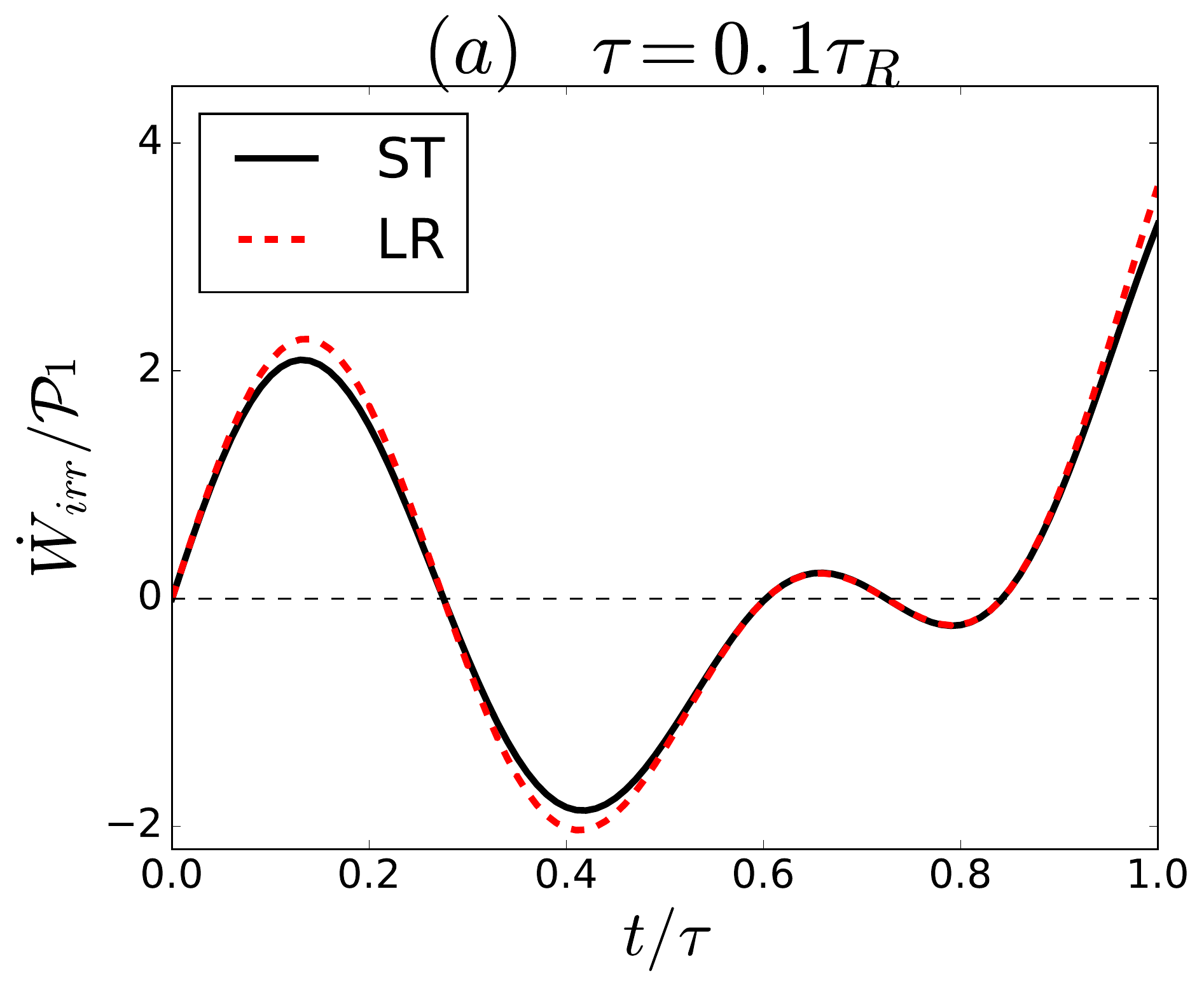}}
    \subfigure{\includegraphics[scale=0.42]{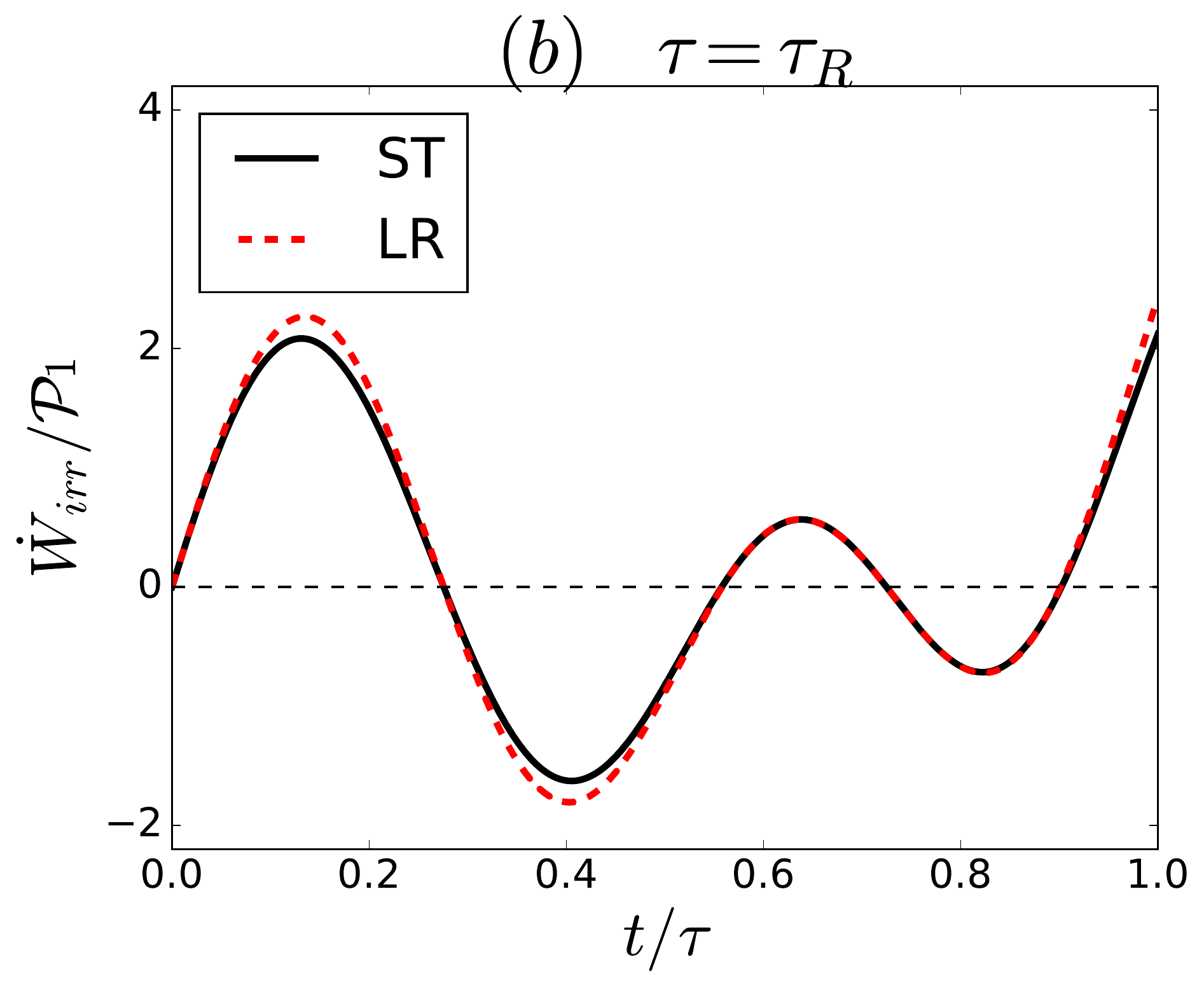}}\\
    \subfigure{\includegraphics[scale=0.42]{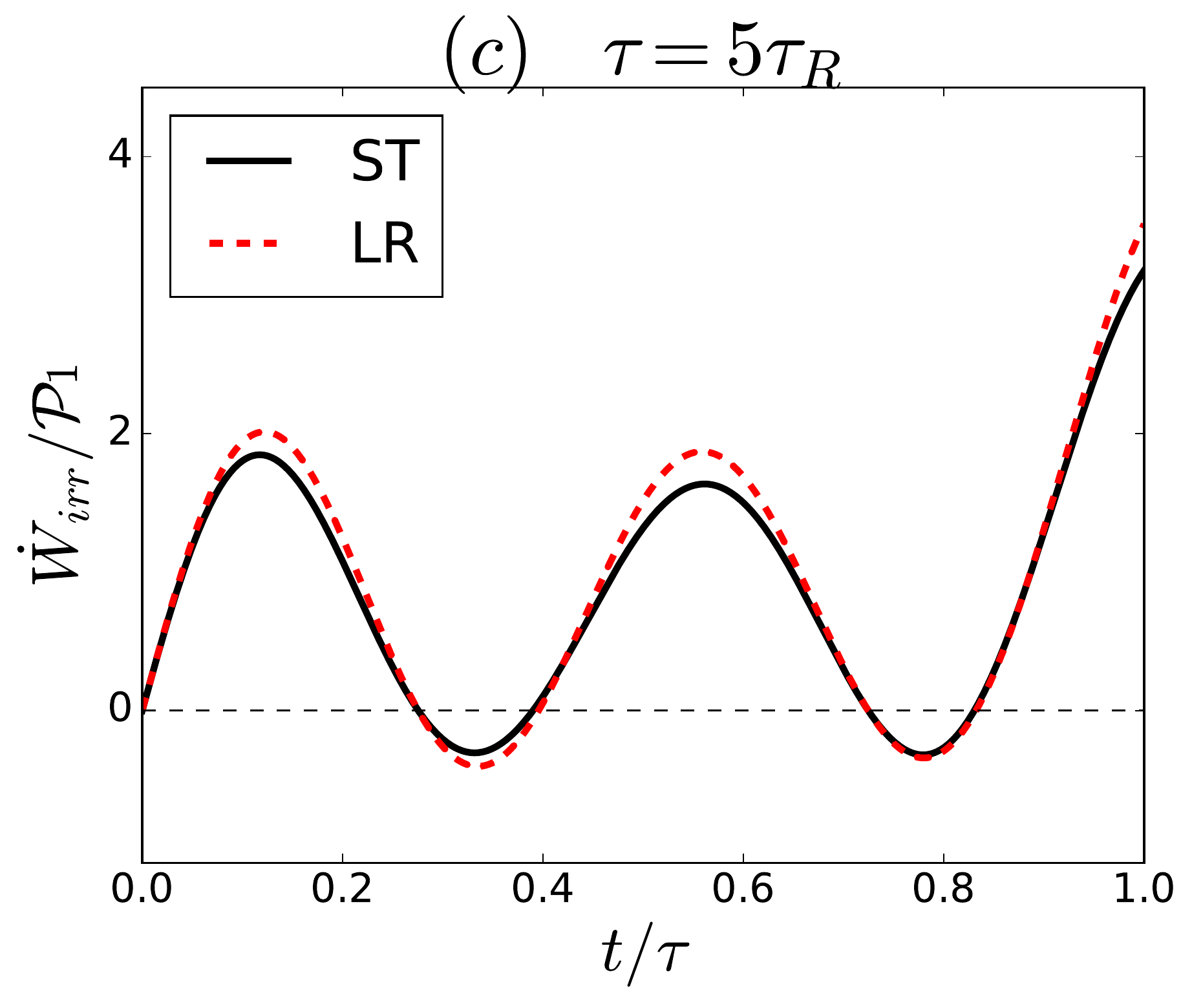}}
    \subfigure{\includegraphics[scale=0.42]{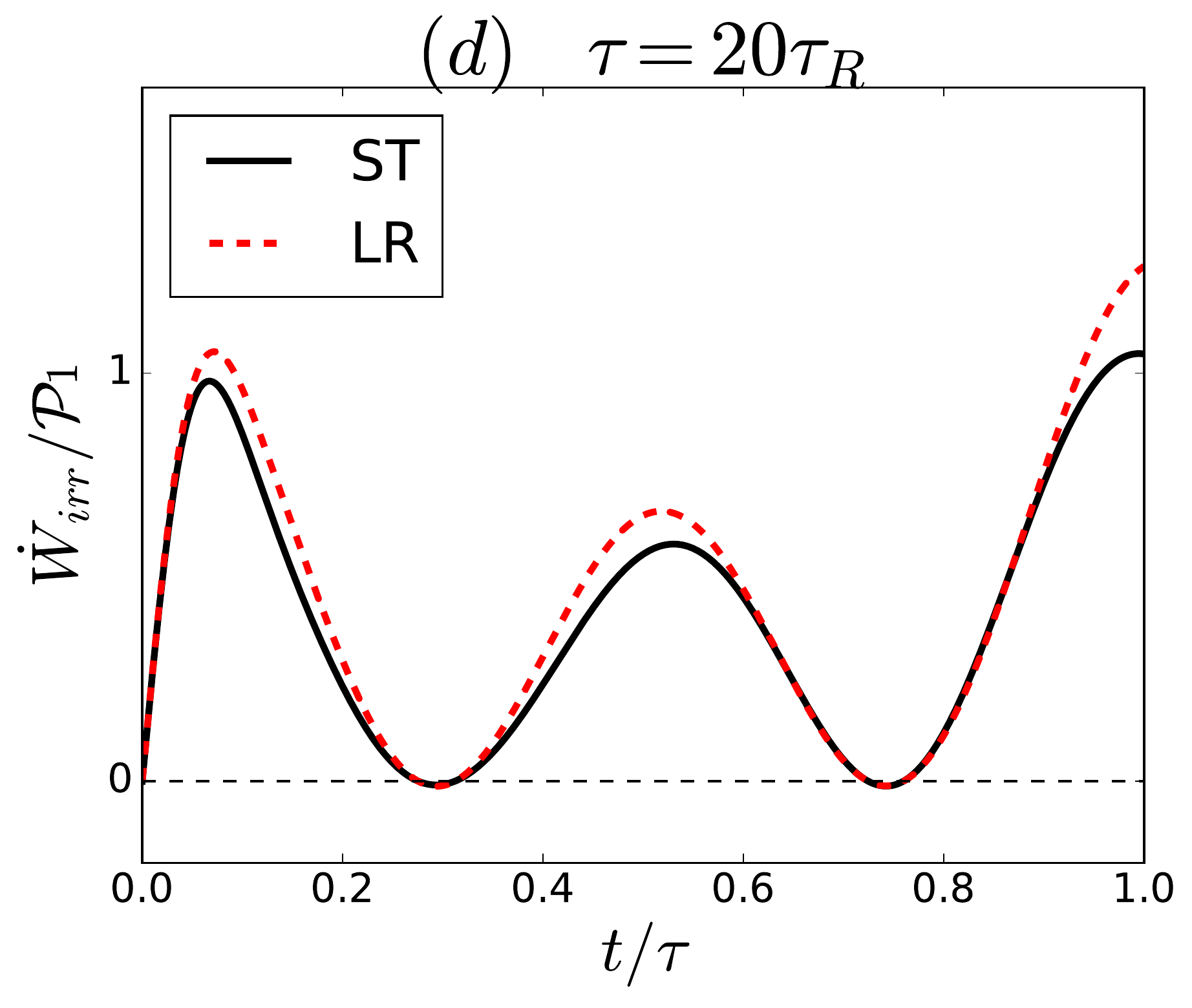}}
    \caption{(Color online) Comparison between the entropy production rate calculated by stochastic thermodynamics (ST) and linear-response theory (LR), using respectively Eqs. (\ref{eq:ipst}) and (\ref{eq:iplr}) with the protocol (\ref{eq:protocol}). As illustrated in panels (a) to (d), negative values of the entropy production rate vanish as the process approaches the quasistatic regime, i.e., the ratio $\tau/\tau_{R}$ increases. We chose $\tau_R=1$, $\gamma/\omega_0=1$, $\delta\lambda/\lambda_0=0.1$ and $m=1$. We chose also the instantaneous power unit as $\mathcal{P}_1 = k_B T/(100\tau)$.}
\label{fig:EPR}
\end{figure*}

\begin{figure}
\includegraphics[scale=0.42]{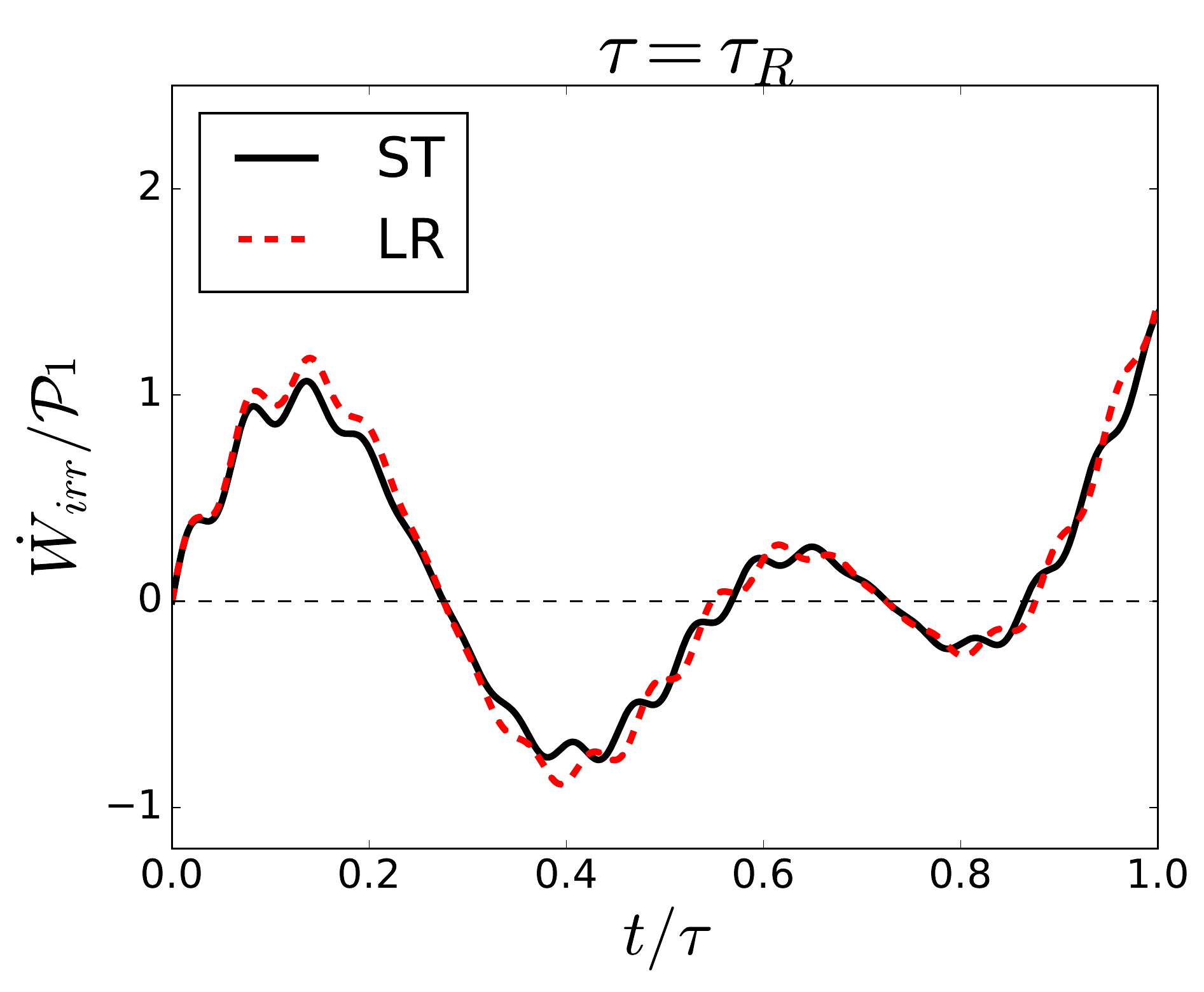}
\caption{(Color online) Comparison between the entropy production rate calculated by stochastic thermodynamics (ST) and linear-response theory (LR), using respectively Eqs. (\ref{eq:ipst}) and (\ref{eq:iplr}) with the protocol (\ref{eq:protocol}). The weak coupling between the system and heat bath does not affect the emergence of negative entropy production rate, although the outlines of the curves are more oscillatory than those of Fig.~\ref{fig:EPR}. We chose $\tau_R=1$, $\gamma/\omega_0=0.01$, $\delta\lambda/\lambda_0=0.1$ and $m=1$. We chose also the instantaneous power unit as $\mathcal{P}_1 = k_B T/(100\tau)$.}
\label{fig:EPR_WC}
\end{figure}

Figure $\ref{fig:EPR}$ shows the comparison between the results obtained using stochastic thermodynamics (ST) and linear-response theory (LR), i.e., Eqs.~(\ref{eq:ipst}) and (\ref{eq:iplr}). The system of equations (\ref{eq:underdamped}) were solved numerically using Eq.~(\ref{eq:protocol}). As the quasistatic limit is approached, $\tau \gg \tau_R$, the entropy production rate becomes non-negative. Although $\gamma/\omega_{0}=1$ for this set of results, we have not observed any noteworthy changes in the results for other ratios (see, for example, Fig \ref{fig:EPR_WC}). The relative change of the control parameter $\lambda$ was chosen to be $\delta\lambda/\lambda_{0}=0.1$ and $m=1$.

The appearance of negative values in the entropy production rate is closely related to the system's memory in responding to the perturbation and the arbitrary affinity or protocol $\lambda(t)$ that the external agent can choose. We remind that the system's memory manifests itself as a delayed response of the flux in respect to the affinity of the system. Roughly speaking, the convolution between $\Psi_{0}$ and $\dot{\lambda}(t)$ in (\ref{eq:iplr}) gives rise to a flux that is out-of-phase with $\dot{\lambda}(t)$. In addition, the affinity can be positive or negative as long as the protocol $\lambda(t)$ is non-monotonic. The combination of these aspects entails a product of terms with different signs in different regions possibly yielding negative rates. To illustrate this, Eq.~(\ref{eq:iplr}) can be written as the product of a delayed propagation of the affinity,

\be
\text{Prop}_{\Psi_0}[\dot{\lambda}] = \int_{0}^t\Psi_0(t-t')\dot{\lambda}(t')dt',
\label{eq:propdel}
\ee
and an instantaneous one,
\be
\text{Prop}_{\delta}[\dot{\lambda}] = \int_{0}^t\delta(t-t')\dot{\lambda}(t')dt',
\label{eq:propinst}
\ee
yealding
\be
\dot{W}^{LR}_{\text{irr}}(t) = \text{Prop}_{\delta}[\dot{\lambda}]\times \text{Prop}_{\Psi_0}[\dot{\lambda}].
\label{eq:prop}
\ee

\begin{figure}
    \includegraphics[scale=0.42]{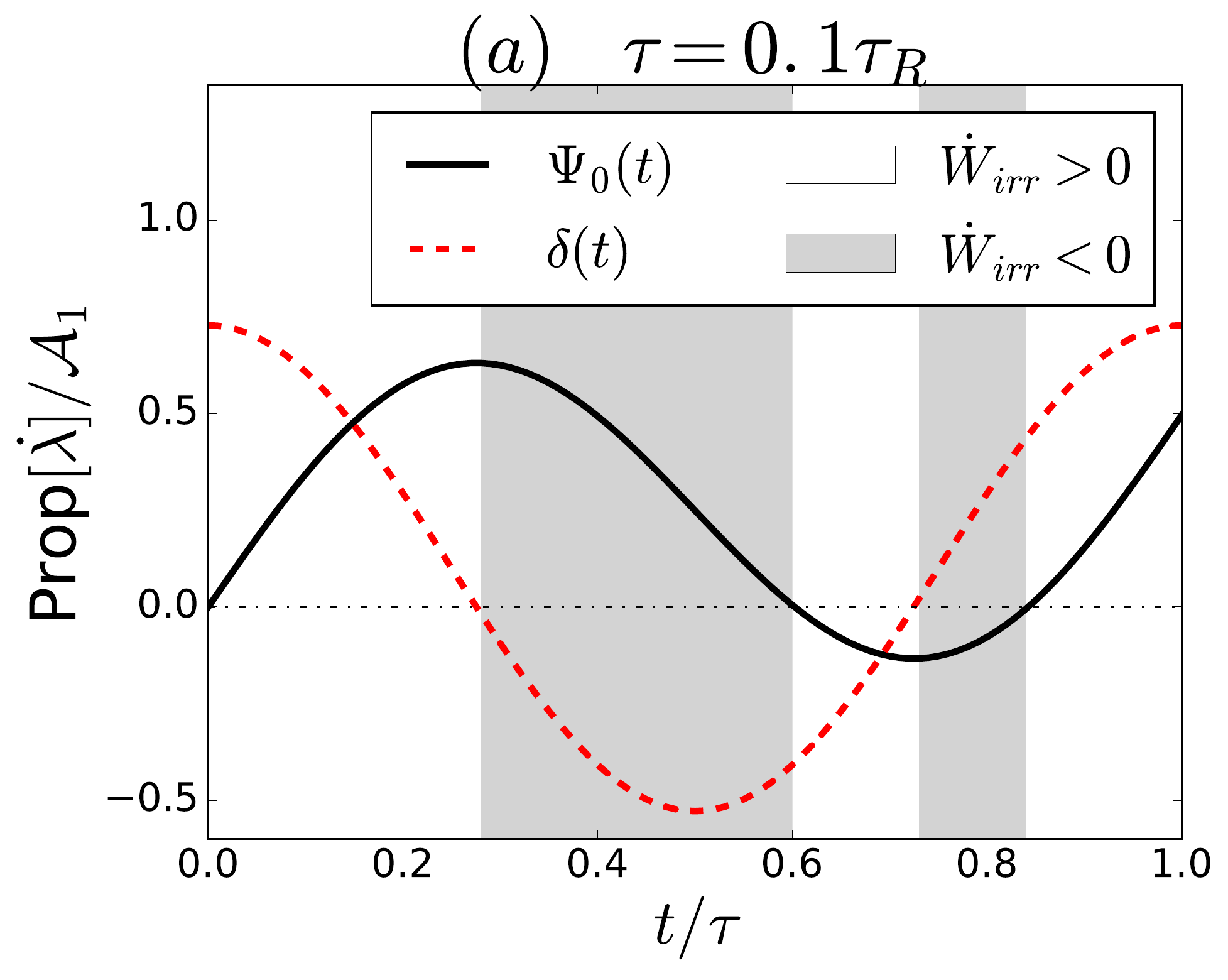}
    \includegraphics[scale=0.42]{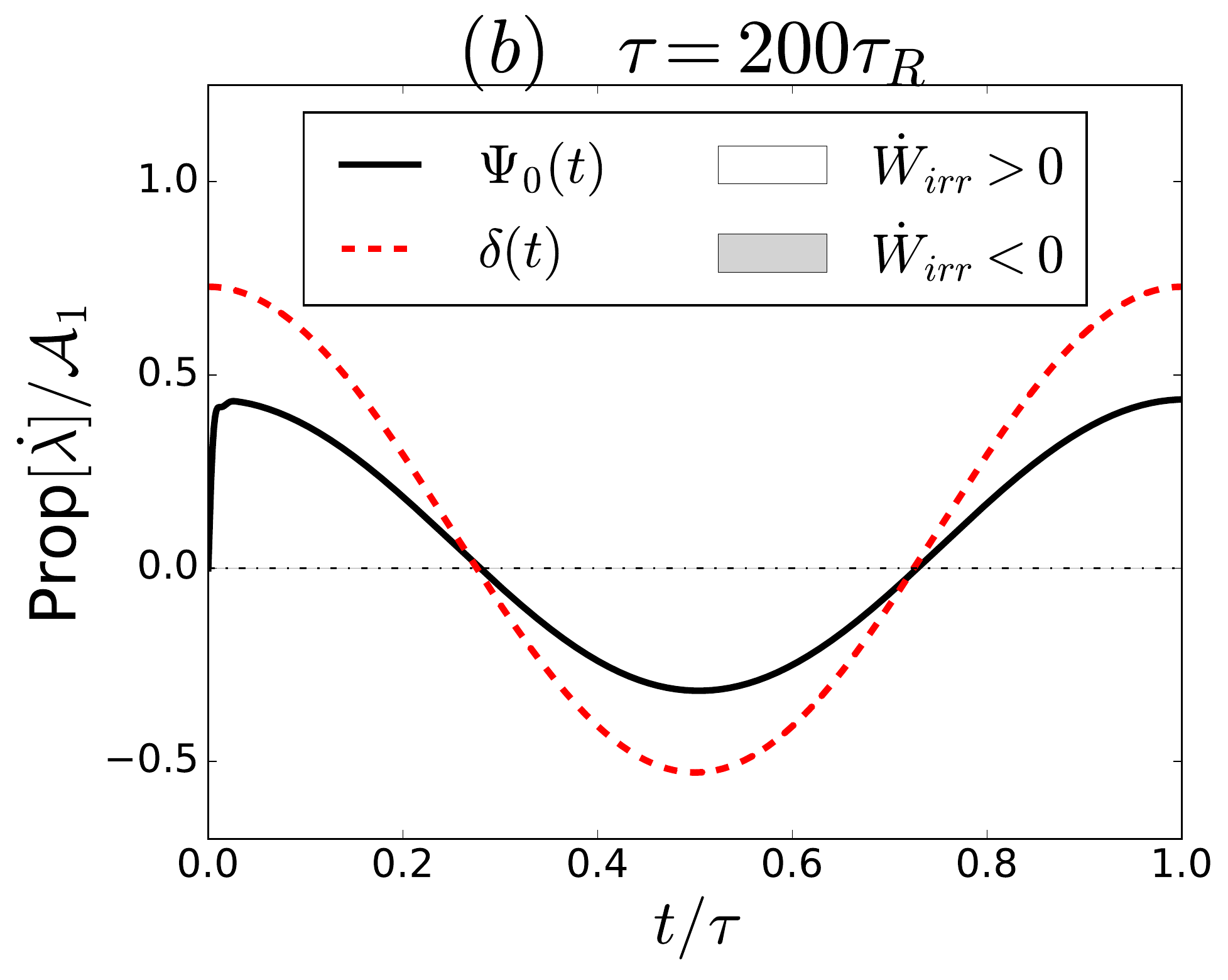}
    \caption{(Color online) Comparisons between the delayed and instantaneous propagations of $\dot{\lambda}$ via the relaxation function and Dirac delta function (see Eqs.~(\ref{eq:propdel}) and (\ref{eq:propinst}) respectively) for the protocol (\ref{eq:protocol}). Panels $(a)$ and $(b)$ illustrate such propagations for $\tau=0.1\tau_R$ and $\tau=200\tau_R$. In (a), the negative values in the entropy production rate are a consequence of the system's memory and the non-monotonicity of the protocol. In (b), the positivity is acquired as long as the response of the flux in respect to the affinity becomes instantaneous. We chose $\tau_R=1$, $\gamma/\omega_0=1$, $\delta\lambda/\lambda_0=0.1$ and $m=1$. We chose also the affinity unit as $\mathcal{A}_1 = \delta\lambda/(10\tau)$. We remark that the delayed propagations $\text{Prop}_{\Psi_0}[\dot{\lambda}]$ were rescaled by factors $(a)$100 and $(b)$1.2 for a better presentation.}
\label{fig:prop}
\end{figure}

Figure \ref{fig:prop} shows the functions obtained from the propagations in Eqs.~(\ref{eq:propdel}) and (\ref{eq:propinst}) for a fast and slow realization of protocol (\ref{eq:protocol}). The positivity of the entropy production rate for the slow process is quite understandable if we remind ourselves of the discussion in Sec. \ref{sec:csvp}. In fact, when $\tau/\tau_{R}\gg 1$, the system will almost relax completely at each small piece of the process. In other words, the process will be entering the slowly-varying regime (see in Fig. \ref{fig:diagram_noneq} the intersection of regions 1 and 2). Therefore, the entropy production rate will become a positive function. From a mathematical point of view, such attainment of the quasistatic regime can be understood as the limit in which the relaxation function becomes a Dirac delta function and the response of the associated flux is basically instantaneous (see Fig. \ref{fig:prop}$(b)$).

\begin{figure}
    \subfigure{\includegraphics[scale=0.42]{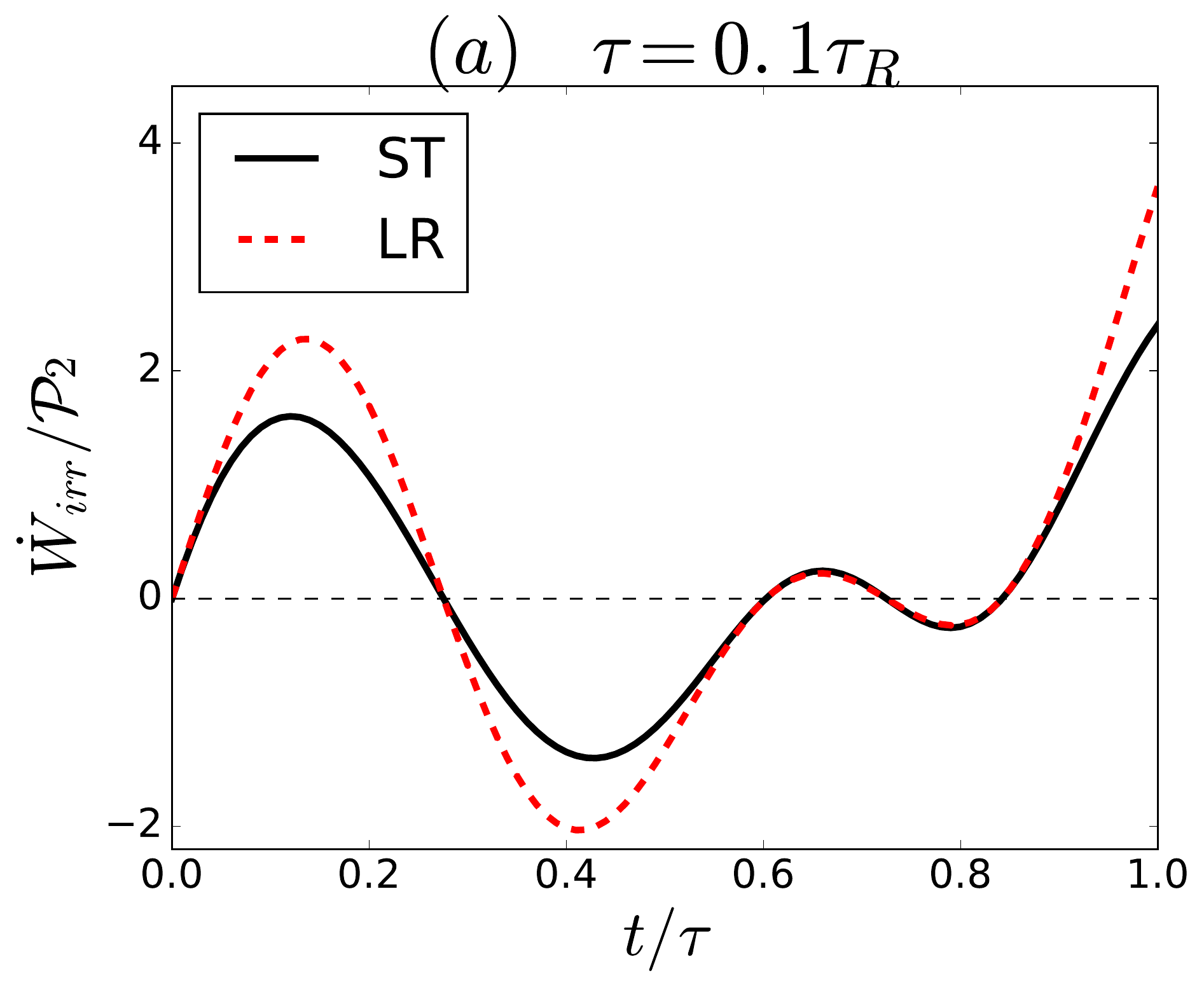}}
    \caption{(Color online) Comparison between the entropy production rate calculated by stochastic thermodynamics (ST) and linear-response theory (LR), using respectively Eqs. (\ref{eq:ipst}) and (\ref{eq:iplr}) with the protocol (\ref{eq:protocol}). We observe that LR result has the same outline and order of magnitude of ST result. We chose $\tau_R=1$, $\gamma/\omega_0=1$, $\delta\lambda/\lambda_0=0.5$ and $m=1$. We chose also the instantaneous power unit as $\mathcal{P}_2 = k_B T/(4\tau)$.}
\label{fig:EPR150}
\end{figure}

Figure \ref{fig:EPR150} shows a comparison between the entropy production rates calculated by Eqs. (\ref{eq:ipst}) and (\ref{eq:iplr}) with $\delta\lambda/\lambda_{0}=0.5$. Although this is a regime in which LR already deviates from the exact result, it still predicts correctly the time intervals in which negative rates exist. Additionally, the order of magnitude and the outline of the entropy production rate also follow reasonably well the exact result. This shows how useful LR predictions can be even in the fully nonequilibrium regime.

\begin{figure}
    \includegraphics[scale=0.42]{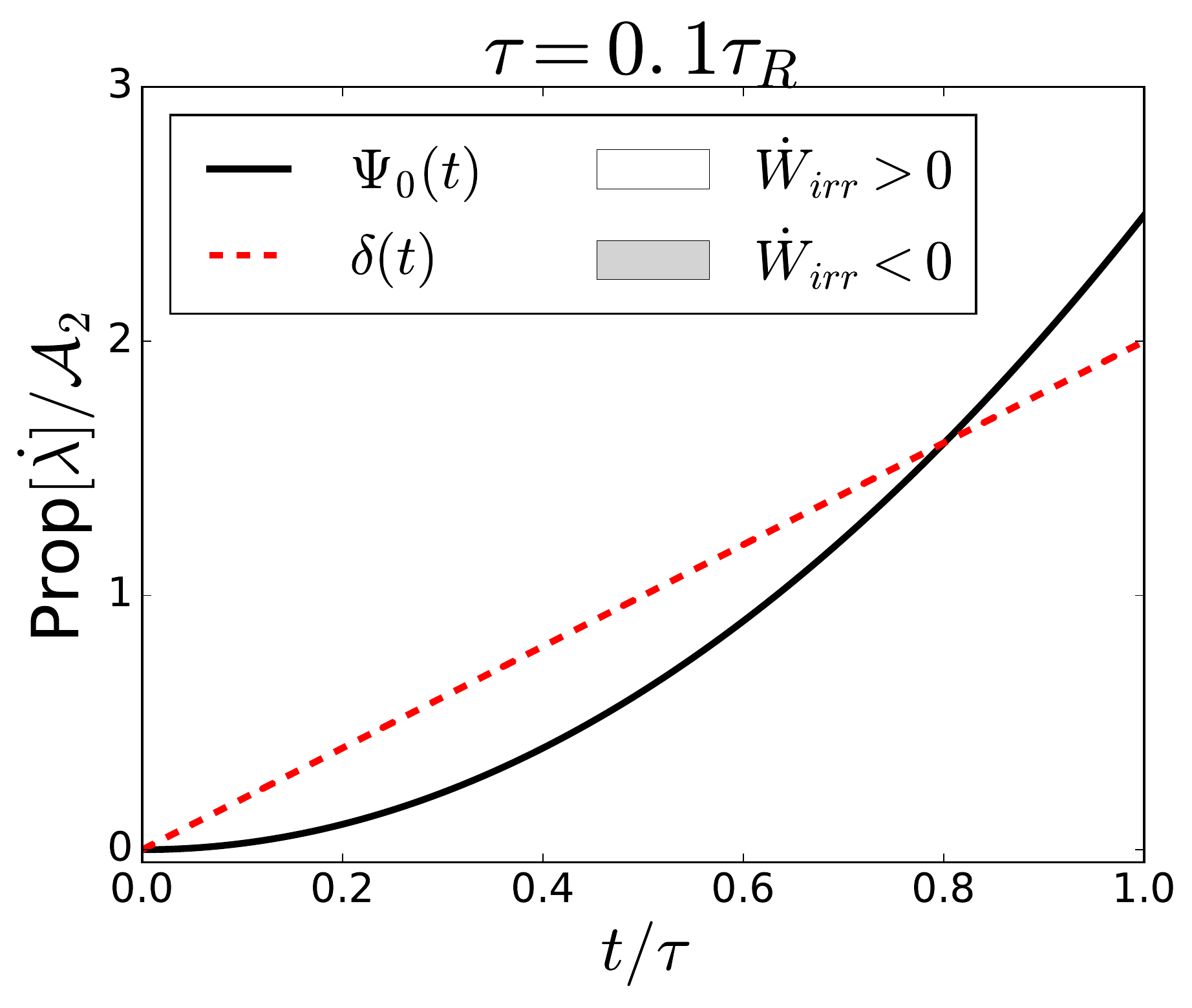}
    \caption{(Color online) Comparison between the delayed and instantaneous propagations of $\dot{\lambda}$ via the relaxation function, given by Eq.~(\ref{eq:relax2}), and Dirac delta function (see Eqs.~(\ref{eq:propdel}) and (\ref{eq:propinst}) respectively). We used the quadratic protocol $g(t)=(t/\tau)^2$. The entropy production rate is positive although the protocol is fast, $\tau=0.1\tau_R$. We chose $\tau_R=1$, $\gamma/\omega_0=1$, $\delta\lambda/\lambda_0=0.1$ and $m=1$. We chose also the affinity unit as $\mathcal{A}_2 = \delta\lambda/\tau$. We remark that the delayed propagation $\text{Prop}_{\Psi_0}[\dot{\lambda}]$ was rescaled by a factor 50 for a better presentation.}
\label{fig:EPRlinear}
\end{figure} 

\begin{figure}
	\includegraphics[scale=0.42]{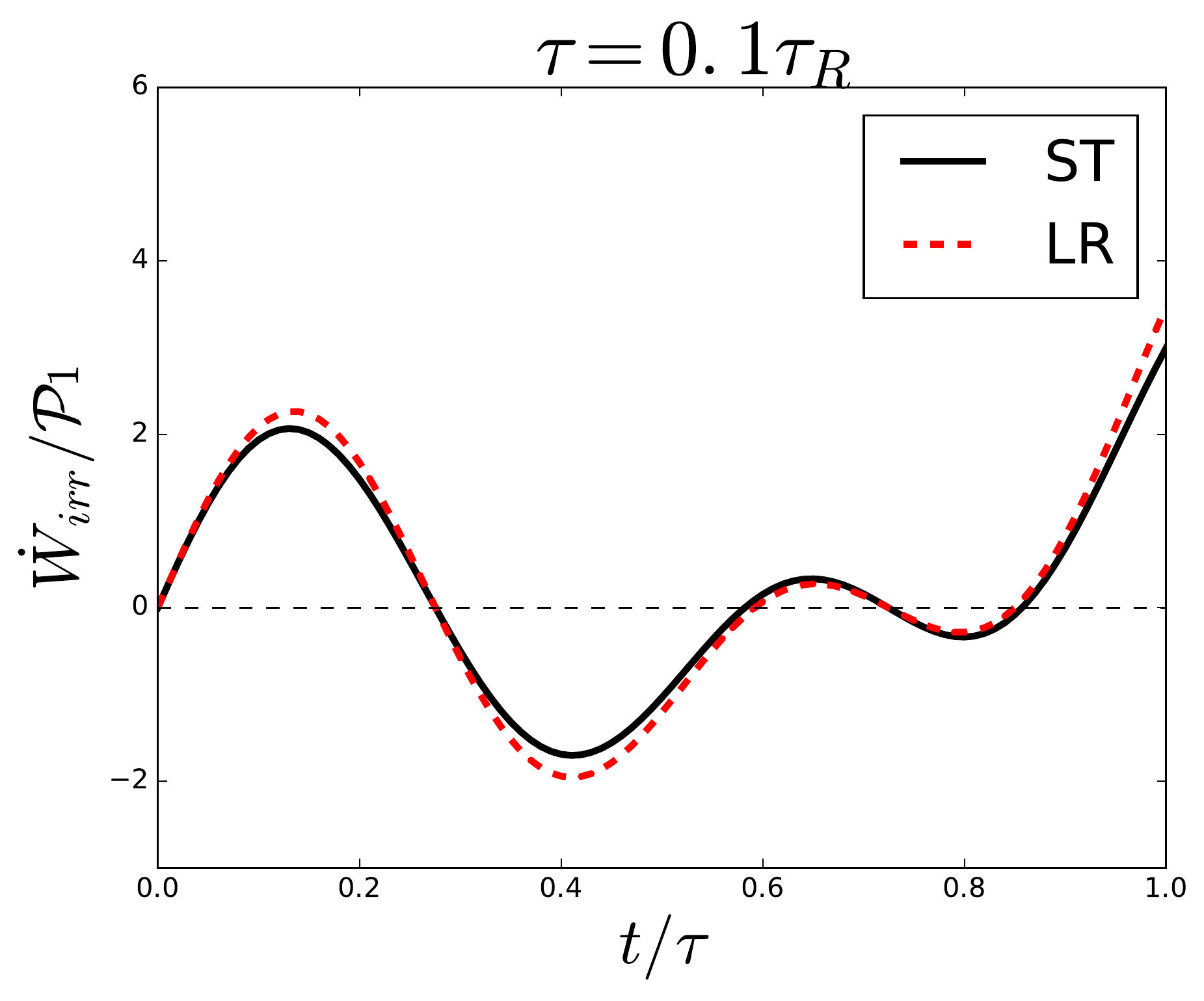}
    \caption{(Color online) Entropy production rates calculated via stochastic thermodynamics (ST) and linear-response theory (LR) using protocol (\ref{eq:protocol}) for overdamped Brownian motion. For the ST calculation, we used the solution of Eq.~(\ref{eq:overdamped}) and for LR the relaxation function $\Psi_{0}(t) = \Psi_{0}(0) e^{-t/\tau_{R}}$, where $\Psi_{0}(0)=\beta\langle x^4_0\rangle_0/4$. We chose $\tau_R=1/\gamma=1$ and $\delta\lambda/\lambda_0=0.1$. The instantaneous power unit was chosen as $\mathcal{P}_1 = k_B T/(100\tau).$}
\label{fig:EPR2v}
\end{figure}

We emphasize that the existence of the negative values observed in the entropy production rate is not an exclusive consequence of the memory of the system. For instance, if we consider a monotonic protocol such as $\lambda(t)=\lambda_0+\delta\lambda(t/\tau)^{2}$, whose affinity is $\dot{\lambda}=2t\delta\lambda/\tau^2$, although the process can be fast the entropy production rate is always positive (see Fig.~\ref{fig:EPRlinear}). As pointed out previously, the form of $\lambda(t)$ is essential in obtaining negative rates. On the other hand, if we consider different systems described by relaxation functions such as $\Psi_{1}(t)$, $\Psi_{3}(t)$ and $\Psi_{4}(t)$ and maintain the protocol (\ref{eq:protocol}), we still observe negative values in the entropy production rate (see Fig.~\ref{fig:EPR2v}). For instance, if we consider Brownian motion in the overdamped regime, expression (\ref{eq:relaxfunc}) yields a simple exponential as the phenomenological model for the relaxation function. The instantaneous power (\ref{eq:ipst}) is obtained from the solution of the equation below \cite{seifert2007}
\be
\partial_t\overline{\overline{x^2}} = -2 \lambda(t)\overline{\overline{x^2}}/\gamma+2 k_B T/\gamma,
\label{eq:overdamped}
\ee
and the instantaneous power provided by linear-response theory in Eq.~(\ref{eq:iplr}) can be calculated using $\Psi_{1}(t)$ \cite{deffner2014, bonanca2018}. Figure \ref{fig:EPR2v} shows a comparison of both results with great agreement. Surprisingly, we still observe negative values in the entropy production rate for $\tau_R =1/\gamma=1$ and $\delta\lambda/\lambda_0=0.1$.

We would like to stress that the results of Figs.~\ref{fig:EPR}, \ref{fig:EPR_WC}, \ref{fig:EPR150} and \ref{fig:EPR2v} should not be misunderstood. The regions with negative values do not lead to violations of the inequality (\ref{eq:diss}) since an integration of $\dot{W}_{\mathrm{irr}}$ over an interval of $t/\tau$ with negative values do not give the $W_{\mathrm{irr}}$ for a process that drives the system between the corresponding values of $\lambda$ in that interval.

\section{Final remarks}
\label{sec:final}

Three complementary aspects, often taken as statements of the Second Law of Thermodynamics, were considered in this work in the context of isothermal processes: no dissipation for quasistatic process, dissipation for finite-time processes and positive entropy production rate. We have shown that the linear-response formulation of slowly-varying processes satisfies almost automatically all of them, while for finite-time but weak processes more detailed properties of the relaxation function are demanded, namely, finite relaxation time, time-reversal symmetry and positive Fourier transform. Concerning the entropy production rate, we have shown a simple example in which negative values are present. Surprisingly, these results were corroborated by exact calculations. The linear-response formulation has also proven useful in understanding the origin of such phenomenon. The absence of time-scale separation between the performed process and the relaxation of the system and the non-monotonicity of the protocol chosen by the external agent are essential ingredients for the emergence of negative rates according to this formulation. Negative entropy production rates have been observed for the same model of a Brownian particle considered here in Ref. \cite{esposito2018} although using a different measure of entropy production in terms of a nonequilibirum free energy.

The clear contrast between our results and previous analysis of the entropy production rates for driven Brownian motion deserves further investigation. In Ref.~\cite{seifert2005}, the non-negativity of the entropy production rate is obtain after considering the ensemble average over microscopic, or trajectory dependent, entropy production rates. Here, we have verified that the rate of a macroscopically-motivated definition of entropy production namely, the irreversible work, can be negative. A careful analysis of the differences between this two quantities will be done in a future work. Nevertheless, it is interesting to observe that in either case the time integral of the corresponding rates leads to non-negative results.

In addition, we remark that it is quite interesting how Hamiltonian constraints are important to make the connection with the Second Law of Thermodynamics. If, on one hand, the ``breakdown" of time-reversal symmetry of microscopic laws in macroscopic irreversible phenomena is often stated as a conundrum, our formulation shows that time-reversal symmetry is necessary in order to achieve compatibility with the Second Law. Along the same line of thought, we wonder whether the positivity of the Fourier transform of the relaxation function, required by thermodynamic constraints, might be also related to Hamiltonian dynamics. We also wonder how this is related to the fact that, although there are regions of negative entropy production rates, the integral over the curves in Figs. \ref{fig:EPR} and \ref{fig:EPR2v} must be positive and hence the maximal amount of negative regions must be already encoded in the relaxation function no matter the protocol we choose.

Finally, the introduction of positive entropy production rate as a statement of the Second Law was made under the assumption of local equilibrium
\cite{mazur1962,prigogine1967}. From the point of view adopted in this work, this would correspond to the regime of slowly-varying processes whose duration is longer than the relaxation time. Once the ratio between relaxation and switching time increases, we enter the region where local equilibrium does not hold anymore and hence we should not expect only positive rates. We consider this result as a significant step towards the understanding of optimal processes in finite time. As in Ref. \cite{seifert2007}, our results suggest that, if we are constrained to switching times of the order of the relaxation time, a non-monotonic protocol does a better job in minimizing dissipation and entropy production. We leave for future work the extensions of our results to more than one control parameter and to the quantum regime.

\section*{Acknowledgements}

The authors thank Sebastian Deffner and Philipp Strasberg for very useful discussions. The authors acknowledge support from FAPESP (Funda\c{c}\~{a}o de Amparo \`a Pesquisa do Estado de S\~ao Paulo) (Brazil) (Grant No. 2018/06365-4 and No. 2018/21285-7) and support from CNPq (Conselho Nacional de Desenvolvimento Cient\'ifico e Pesquisa) (Brazil) (Grant No. 141018/2017-8). 

\appendix
\section{Linear-response expression for the generalized force}
\label{app:a}

In this appendix we provide a quick derivation of Eq.~(\ref{eq:genforce}) for the generalized force. The general idea of the method is to expand the quantities involved in the calculation in terms of $\delta\lambda$ which is assumed to be small. Consider then the non-equilibrium average of $\partial_{\lambda}\mathcal{H}$, given by
\be
\overline{\partial_{\lambda}\mathcal{H}}(t) = \int_\Gamma d\Gamma \rho(\Gamma,t)\,\partial_{\lambda}\mathcal{H}(\Gamma,\lambda(t)),
\label{eq:a1}
\ee
where $\rho$ is the nonequilibrium ensemble of the total system and $\Gamma$ is a point in the phase space. Firstly, the expansion of the Hamiltonian $\mathcal{H}(\Gamma,\lambda(t))$ in the perturbation $\delta\lambda$ reads
\be
\mathcal{H}(\Gamma,\lambda(t)) = \mathcal{H}(\Gamma,\lambda_0)+\partial_{\lambda}\mathcal{H}(\Gamma,\lambda_0)g(t)\delta\lambda+\mathcal{O}(\delta\lambda^2),
\label{eq:a2}
\ee
and, consequently,
\be
\partial_{\lambda}\mathcal{H}(\Gamma,\lambda(t)) = \partial_{\lambda}\mathcal{H}(\Gamma,\lambda_0)+\partial^2_{\lambda\lambda}\mathcal{H}(\Gamma,\lambda_0)g(t)\delta\lambda+\mathcal{O}(\delta\lambda^2).
\label{eq:a3}
\ee
On the other hand, the non-equilibrium ensemble $\rho$ must satisfy the Liouville equation
\be
\partial_t\rho = -\{\rho,\mathcal{H}\} := \mathcal{L}\rho,
\label{eq:a4}
\ee
where $\mathcal{L}$ is called the Liouville operator. Equation~(\ref{eq:a4}) can be written in the following integral form,
\be
\rho(\Gamma,t) = \rho_{\text{eq}}(\Gamma)+\int_{0}^{t}e^{-\mathcal{L}(t-t')}\mathcal{L}\rho(\Gamma,t')dt',
\label{eq:a5}
\ee
where $e^{-\mathcal{L}t'}$ is the dynamical evolution operator of the ensemble $\rho(\Gamma,0)$. Using Eq.~(\ref{eq:a2}) and (\ref{eq:a5}), the non-equilibrium ensemble $\rho$ expanded until its first order in $\delta\lambda$ is given by
\be
\rho(\Gamma,t) = \rho_{\text{eq}}(\Gamma)+ \int_{0}^{t}e^{-\mathcal{L}_0(t-t')}\mathcal{L}_1\rho_{\text{eq}}(\Gamma)dt'+\mathcal{O}(\delta\lambda^2),
\label{eq:a6}
\ee
where we defined the following Liouville operators
\be
\mathcal{L}_0(\cdot) := -\{\cdot,\mathcal{H}_0\},\quad \mathcal{L}_1(\cdot) := -\{\cdot,\partial_{\lambda}\mathcal{H}(\Gamma,\lambda_0)\}g(t)\delta\lambda.
\label{eq:a7}
\ee
Using Eqs.~(\ref{eq:a1}), (\ref{eq:a3}) and (\ref{eq:a6}), the anti-Hermiticity property of the Liouville operators and restraining us to the first order in $\delta\lambda$, we obtain
\begin{equation}
\begin{split}
\overline{\partial_{\lambda}\mathcal{H}}(t) &= \langle\partial_{\lambda}\mathcal{H}(\Gamma,\lambda_0)\rangle_0+\langle\partial^2_{\lambda\lambda}\mathcal{H}(\Gamma,\lambda_0)\rangle_0g(t)\delta\lambda\\
&\quad+\delta\lambda\int_0^t dt'\Phi_0(t-t')g(t'), 
\label{eq:a8}
\end{split}
\end{equation}
where $\langle A\rangle_{0}$ denotes the average of the observable $A$ taken with $\rho_{\text{eq}}$,
\be
\Phi_0(t) = \langle\{\partial_{\lambda}\mathcal{H}(\Gamma,\lambda_0),\partial_{\lambda}\mathcal{H}(\Gamma_t,\lambda_0)\}\rangle_{0}
\label{eq:a9}
\ee
is the so-called response function and $\Gamma_t$ is the phase-space point evolved up to time $t$. Defining the relaxation function $\Psi_0(t)$ as
\be
\Psi_0(t) := -\int \Phi_0(t) dt-\mathcal{C}\,,
\label{eq:a10}
\ee
and using Kubo formula \cite{kubo1985} for the canonical distribution $\rho_{\text{eq}}(\Gamma) = \exp{(-\beta \mathcal{H}(\Gamma))}/Z(\beta)$, 
\begin{equation}
\Phi_0(t) = -\beta \frac{d}{dt}\langle \partial_{\lambda}\mathcal{H}(0)\partial_{\lambda}\mathcal{H}(t)\rangle_{0}\,,
\label{eq:a10a}
\end{equation}
one obtains Eq.~(\ref{eq:relax}) in Sec.~\ref{sec:lrt}. Besides, after an integration by parts in Eq.~(\ref{eq:a8}), we obtain
\begin{eqnarray}
\lefteqn{\overline{\partial_\lambda \mathcal{H}}(t) =}\nonumber\\
&& \left\langle\partial_\lambda \mathcal{H}\right\rangle_0-\delta\lambda \Theta_0 g(t) +\delta\lambda\int_{0}^{t} dt'\Psi_0(t')\dot{g}(t-t')+\mathcal{O}(\delta\lambda^2),\nonumber \\
\label{eq:a11}
\end{eqnarray}
where
\be
\Theta_0 := \Psi_0(0)-\left\langle\partial^2_{\lambda\lambda} \mathcal{H}(\Gamma,\lambda_0)\right\rangle_0.
\label{eq:atheta}
\ee
We refer to Refs.~\cite{kubo1985,deffner2014,bonanca2015} for more details.

\section{Linear-response expressions for the irreversible work}
\label{app:b}

We shall start with the irreversible work for slowly varying processes, given by
\be
W_{\text{irr}} = \beta\int_0^\tau\dot{\lambda}^2(t)\tau_R[\lambda(t)]\chi[\lambda(t)]dt.
\label{eq:b1}
\ee
We split now the processes of duration $\tau$ in $N$ time steps of length $\delta t := \tau/N$. Along the $n$-th time step, the control parameter $\lambda$, whose change is described by the protocol $\lambda_n(t) = \lambda_n + \delta\lambda_n g_n(t)$, varies only $\delta\lambda_n$, which is considered to be small enough so that we can use linear-response theory to describe the generalized force. In particular, the generalized force $\overline{\partial_\lambda \mathcal{H}}_n(t)$ for the $n$-th time step is
\begin{eqnarray}
\lefteqn{\overline{\partial_\lambda \mathcal{H}}_n(t) =}\nonumber\\
&& \left\langle\partial_\lambda \mathcal{H}\right\rangle_n-\delta\lambda_n \Theta_n g_n(t) +\delta\lambda_n\dot{g}_n\,\Psi_n(0)\tau_R(\lambda_n),
\label{eq:genforcesv}
\end{eqnarray} 
where the index $n$ indicates that all quantities involved are calculated in the $n$-th time step and the equilibrium averages are taken with $\lambda=\lambda_{n}$ (we are using here the same notation introduced in Appendix~\ref{app:a} for the averages $\overline{A}$ and $\langle A\rangle$ of a given observable $A$). Expression (\ref{eq:genforcesv}) is obtained from Eq.~(\ref{eq:a11}) considering that, in the interval $\delta t$, the protocol $g_n$ can be taken as linear (which implies that $\dot g_n$ is constant but dependent on the $n$-th step), and that the system relaxes faster than $\delta t$, so that we can use the definition of the relaxation time, given by Eq.~(\ref{eq:timerelax}) of Sec.~\ref{sec:csvp}, in Eq.~(\ref{eq:a11}). We denote by $\Psi_{n}(0)$ the amplitude of the relaxation function at the beginning of the $n$-th time step. In other words, the argument ``$0$" in $\Psi_{n}(0)$ refers to the instant of time in which the $n$-th step begins. The constant $\Theta_{n}$ is given by Eq.~(\ref{eq:atheta}) with $\lambda_{n}$ replacing $\lambda_{0}$.

Finally, the work $\delta W_{n}$ performed on the system in the interval $\delta t$ during the $n$-th step reads
\begin{equation}
\begin{split}
\delta W_{n} = \left\langle\partial_\lambda \mathcal{H}\right\rangle_{n}\dot g_n\delta\lambda_n\delta t-\Theta_n \dot{g}_n g_{n}(t)(\delta\lambda_n)^2\delta t\\+(\delta\lambda_n)^2\dot{g}^2_n\Psi_{n}(0)\tau_R(\lambda_{n})\delta t.
\end{split}
\label{eq:b2}
\end{equation}
It can be shown \cite{deffner2014} that the first two terms of the previous expression lead to the free-energy change. Therefore, the irreversible contribution comes from the third term, which has already the form of the integrand in Eq.~(\ref{eq:b1}). Considering that $\Psi_{n}(0) = \beta\chi(\lambda_{n})$, where $\chi$ is the variance of the $\partial_\lambda \mathcal{H}$ evaluated at $\lambda=\lambda_{n}$, and summing up all the irreversible contributions of each $n$-th step,  we obtain in the continuum limit,
\be
W_{\mathrm{irr}} = \beta\int_0^\tau\dot{\lambda}^2(t)\tau_R[\lambda(t)]\chi[\lambda(t)]dt,
\ee
where we consider the $n$-th dependence of the quantities as a instantaneous dependence on the parameter $\lambda(t)$. 

The irreversible work expression for finite-time but weak processes comes up from the very definition of work,
\begin{equation}
W(\tau) = \int_0^\tau dt\,\dot{\lambda}(t)\,\overline{\partial_{\lambda}\mathcal{H}}(t)\,,
\label{eq:workdef}
\end{equation}
and the linear-response expression we have already derived in Appendix~\ref{app:a} for the generalized force,
\begin{eqnarray}
\lefteqn{\overline{\partial_\lambda \mathcal{H}}(t) =}\nonumber\\
&& \left\langle\partial_\lambda \mathcal{H}\right\rangle_0-\delta\lambda \Theta_0 g(t) +\delta\lambda\int_{0}^{t} dt'\Psi_0(t')\dot{g}(t-t')+\mathcal{O}(\delta\lambda^2),\nonumber \\
\end{eqnarray}
Plugging the third term of the previous expression into (\ref{eq:workdef}), we obtain the irreversible contribution given by Eq.~(\ref{eq:irrwork}). For more details, see Ref.~\cite{bonanca2015}.

\section{Hamiltonian constraints on response functions}
\label{app:c}
As mentioned in Sec.~\ref{sec:lrt}, Hamiltonian dynamics imposes certain constraints on phenomenological expressions for the response functions. To exemplify what we mean by this, consider the response function $\Phi_{0}(t)$ (see Appendix~\ref{app:a} for more details and definitions),
\be
\Phi_0(t) = \langle\{\partial_{\lambda}\mathcal{H}(0),\partial_{\lambda}\mathcal{H}(t)\}\rangle_0 = -\frac{d\Psi_0(t)}{dt}\,,
\label{eq:responsefunc}
\ee
where $\{.,.\}$ denotes the Poisson bracket and $\langle A\rangle_{0}$ denotes an equilibrium average of observable $A$. The short-time expansion of $\Phi_{0}(t)$ reads
\be
\Phi_0(t) = \Phi^{(0)}_0(0)+\Phi^{(1)}_0(0) t+\Phi^{(2)}_0(0) \frac{t^2}{2}+...,
\ee
where the coefficients are given by \cite{kubo1985,kubo1972}
\begin{subequations}\label{eq:respocoef}
\begin{align}
\Phi_{0}^{(0)}(0) &= \langle \{\partial_{\lambda}\mathcal{H}(0),\partial_{\lambda}\mathcal{H}(0)\}\rangle_0 = 0, \\
\Phi_{0}^{(1)}(0) &= \langle \{\partial_{\lambda}\mathcal{H}(0),\{\partial_{\lambda}\mathcal{H}(0),\mathcal{H}\}\}\rangle_0, \\
\Phi_{0}^{(2)}(0) &= \langle \{\partial_{\lambda}\mathcal{H}(0),\{\{\partial_{\lambda}\mathcal{H}(0),\mathcal{H}\},\mathcal{H}\}\}\rangle_0=0.
\end{align}
\end{subequations}

Indeed, the Hamiltonian dynamics demand $\Phi_{0}(t)$ to be \emph{odd} with respect to the change $t\rightarrow -t$ and therefore all $\Phi^{(k)}_{0}(0)$ with $k$ even are zero. Consequently, due to Eq.(\ref{eq:responsefunc}), any phenomenological model of the relaxation function $\Psi_0(t)$ must be \emph{even} under the change $t\rightarrow -t$. Each of the Eqs.~(\ref{eq:respocoef}) are examples of Hamiltonian constraints leading to different sum-rules. These constraints must be imposed to any expression intended to be used as a valid relaxation or response function. Since there is a infinite hierarchy of them, a phenomenological expression with a finite number of free parameters can only fulfill a finite amount of them (for more details of this method, see Ref.~\cite{kubo1972}).

\end{document}